\documentclass[twocolumn,apjthe,colorlinks,linkcolor=blue,citecolor=blue,urlcolor=blue,twocolappendix,numberedappendix]{openjournal}
\usepackage[T1]{fontenc}
\usepackage{orcidlink}
\usepackage{graphicx}	
\graphicspath{{./}{fig/}}
\usepackage{amsmath}	
\usepackage{bm}	
\usepackage{amsbsy}	
\usepackage{xcolor}
\usepackage{multirow}

\newcommand{\lya}{Ly~$\alpha$}

\newcommand{\msolh}{\ensuremath{h^{-1}\,{\rm M}_\odot}}
\newcommand{\msol}{\ensuremath{{\rm M}_\odot}}

\newcommand{\cmpch}{\ensuremath{h^{-1}\,{\rm cMpc}}}

\newcommand{\ud}{{\rm d}}
\newcommand{\angstrom}{\text{\normalfont\AA}}
\newcommand{\thesan}{{\sc thesan}}

\newcommand{\ergs}{\ensuremath{{\rm erg\,s^{-1}}}}
\newcommand{\tauf}{\ensuremath{\tau_{\rm eff}^{50}}}
\newcommand{\Sigmaoiii}{\ensuremath{\Sigma_{\rm \oiii}(r_\perp < 10~\cmpch)}}
\newcommand{\Sigmaoiiim}{\ensuremath{\langle \Sigma_{\rm \oiii} \rangle}}
\newcommand{\lmh}{$L_{\rm UV}(M_h)$}
\newcommand{\eg}{e.g.\ }
\newcommand{\ie}{i.e.\ }
\newcommand{\hi}{H\,\textsc{i}}

\newcommand{\oiii}{[O\,\textsc{iii}]}
\newcommand{\oiiia}{\ensuremath{{\text{[O\,\textsc{iii}]}\lambda5008}}}
\newcommand{\oiiib}{\ensuremath{{\text{[O\,\textsc{iii}]}\lambda4960}}}
\newcommand{\oiiiab}{\ensuremath{{\text{[O\,\textsc{iii}]}\lambda\lambda4960,5008}}}
\newcommand{\loiii}{\ensuremath{L_{\rm [O\,\textsc{iii}]}}}

\newcommand{\eref}[1]{Eq.~\eqref{#1}}
\newcommand{\sref}[1]{Sec.~\ref{#1}}
\newcommand{\fref}[1]{Fig.~\ref{#1}}

\defcitealias{bouwens2021}{B21}
\defcitealias{kakiichi2025}{K25}
\defcitealias{matthee2023}{M23}
\defcitealias{zhu2026}{Z26}

\begin{document}
\title{The impact of source and survey modelling on the connection between \oiii{} emitters and \lya{} forest transmission at $\MakeLowercase{z} \sim 6$}
\shorttitle{\oiii{} emitter source and survey modelling}

\author{Luke Conaboy$^{1,\star}$\,\orcidlink{0000-0002-6580-7177},
  James S. Bolton$^{1,\dag}$\,\orcidlink{0000-0003-2764-8248},
  Laura C. Keating$^{2}$\,\orcidlink{0000-0001-5211-1958}, 
  Martin G. Haehnelt$^{3}$\,\orcidlink{0000-0001-8443-2393}, 
  Girish Kulkarni$^{4}$\,\orcidlink{0000-0001-5829-4716},
  Ewald Puchwein$^{5}$\,\orcidlink{0000-0001-8778-7587}}
\shortauthors{Conaboy et al.}
\thanks{\raggedright E-mail: $^\star${\tt lukeconaboy.astro@gmail.com}, $^\dag${\tt james.bolton@nottingham.ac.uk}}

\affiliation{$^{1}$School of Physics and Astronomy, The University of Nottingham, University Park, Nottingham, NG7 2RD, UK}
\affiliation{$^{2}$Institute for Astronomy, University of Edinburgh, Blackford Hill, Edinburgh, EH9 3HJ, UK}
\affiliation{$^{3}$Kavli Institute for Cosmology and Institute of Astronomy, Madingley Road, Cambridge, CB3 0HA, UK}
\affiliation{$^{4}$Tata Institute of Fundamental Research, Homi Bhabha Road, Mumbai 400005, India}
\affiliation{$^{5}$Leibniz-Institut f\"ur Astrophysik Potsdam, An der Sternwarte 16, 14482 Potsdam, Germany}

\begin{abstract} {\em James Webb Space Telescope} ({\em JWST}) surveys
  of \oiii{}-emitting galaxies are offering fresh insight into the
  connection between galaxies and the intergalactic medium at redshift
  $z\sim 6$. Recent measurements of the cross-correlation between
  \oiii{}-emitting galaxies and \lya{} forest transmission point to
  excess \lya{} transmission at a scale ($r\approx 30~{\rm cMpc}$) and
  amplitude (roughly twice the global mean) that is not seen in
  numerical models. Here we improve upon previous theoretical work by
  constructing an empirical model that connects haloes with the
  observed population of \oiii{} emitters and incorporates the
  geometry and depth of the {\em JWST} surveys into mock galaxy survey
  catalogues. We compare these mocks to recent measurements of \oiii{}
  emitter clustering, the one-dimensional galaxy--\lya{} transmission
  cross-correlation, and the relation between galaxy surface density
  and \lya{} effective optical depth. The large scatter in our mock
  survey measurements of the cross-correlation means there is no
  significant tension with the observational data, albeit the peak of
  the one dimensional correlation in our mocks occurs at a scale
  $\approx 10\rm\,cMpc$ below that observed. Though further studies of the
  effect of different ionising source models will be useful, the large
  scatter implies that, at present, current galaxy-IGM observations
  may struggle to rule out a broad range of such models. We anticipate
  that further progress will strongly benefit from increased
  observational sample sizes, as well as simulations performed in box
  sizes $>250\rm\,cMpc$ that use a variety of source models.
\end{abstract}


\section{Introduction}
\label{sec:introduction}
In the first billion years, the diffuse hydrogen gas between galaxies
transitioned from a neutral to an ionised state -- a process called
reionisation. Observations of the cosmic microwave background suggest
the mid-point of reionisation occurred around redshift $z \approx 7.5$
\citep{planckcollaboration2020, gorce2022}, although this only gives
an integral constraint on the timing. A wide range of other
observables suggest reionisation finished at $z<6$, and perhaps even
as late as $z \approx 5.3$: the distribution of the Lyman~$\alpha$ (\lya{})
forest effective optical depth \citep{bosman2022}; the rapid evolution
of the mean free path of Lyman limit photons around $z \sim 6$
\citep{becker2021, zhu2023, satyavolu2024}; observations of damping
wings associated with \lya{} troughs in the spectra of
$z \lesssim 6$ quasars \citep{becker2024, spina2024, zhu2024}; and the
incidence of dark gaps in the \lya{} forest \citep{mcgreer2015,
  jin2023, davies2026}. However, while there has been great progress
in constraining the timing of the end of reionisation, the nature of
the ionising sources remains an open question.

A promising avenue for gaining insight into the sources driving
reionisation is the correlation of high-redshift galaxies with \lya{}
forest transmission \citep[e.g.][]{kakiichi2018, meyer2019,
  meyer2020}. The advent of {\em JWST} has enabled surveys of
\oiii-emitting galaxies in fields around bright quasars, opening up a
new window into this correlation during the second half of
reionisation (\citealt{kashino2023}, \citealt{kakiichi2025},
hereafter \citetalias{kakiichi2025}, \citealt{kashino2026},
\citealt{zhu2026}, hereafter \citetalias{zhu2026},
\citealt{jin2026}). The \oiii{} line is targeted because it
allows a clear determination of redshift, while also being sensitive
to the ionising photon production of galaxies (\eg
\citealt{wilkins2023}). Large-scale surveys such as FRESCO
\citep{meyer2024} and COSMOS-3D \citep{meyer2025} are targeting the
\oiiiab{} emission line over large areas, while smaller-scale surveys,
like EIGER \citep{kashino2023} and ASPIRE \citep{wang2023},
are focused on the fields of high-redshift quasars. The survey
designs range from single pointings in the case of ASPIRE to more
complex mosaicked tilings in the case of EIGER.

The latest observations of the \oiii{} emitter--\lya\ cross correlation
point to an excess in intergalactic \lya{} transmission relative to
the mean at scales of $\sim 10$--$40~{\rm cMpc}$ from galaxies
(\citetalias{kakiichi2025}, \citealt{kashino2026}). The signature is
driven largely by enhanced ionisation around the galaxies, with
temperature fluctuations due to inhomogeneous reionisation playing a
secondary role \citep{conaboy2025}. A range of theoretical models also
predict this reported excess (\citealt{garaldi2022, garaldi2025a,
  conaboy2025, basu2026}, but see also \citealt{garaldi2019,zhu2024}),
although the observed scale and amplitude are challenging to
reproduce. Recently, \citetalias{zhu2026} have also carried out the
first analysis of the two-dimensional cross-correlation between {\em
  JWST}-detected \oiii{} emitters and \lya\ transmission in the field
of two high-redshift quasars, finding possible evidence for the
anisotropic escape of ionising photons.

However, the modelling of the cross-correlation has not always
considered the detailed properties of the \oiii{} emitter population
or the impact of NIRcam field-of-view and survey depth (with the
recent notable exception of \citealt{garaldi2025a}). In this context,
we extend our earlier work using the Sherwood-Relics simulations
\citep{conaboy2025}, in which the \oiii{} emitters were assigned to
dark matter haloes using a cut in halo mass with
$M_h\geq10^{10}~\msolh$. Using this simple approach, \citet{conaboy2025}
found reasonable quantitative agreement ($< 1.5\sigma$) with the ASPIRE
cross-correlation results presented in \citetalias{kakiichi2025}, but with
a peak in the correlation that occurred at scales $\sim 10 \rm\,cMpc$
below that observed. Here we develop a more detailed empirical model
to connect the haloes in our simulation with the population of \oiii{}
emitters observed by {\em JWST}. We then quantify the impact of this
updated approach on the shape of the galaxy--\lya{} transmission
correlation. Finally, we apply this model to the latest observations
of the \lya{} forest opacity and galaxy density relation using
\oiii{}-selected galaxies \citepalias{zhu2026}.

Throughout this paper we assume a flat $\Lambda$CDM
cosmology with $\Omega_\Lambda=0.692$, $\Omega_{\rm m}=0.308$,
$\Omega_{\rm b}=0.0482$, $\sigma_8=0.829$, $n_s=0.961$ and $h=0.678$
\citep{planckcollaboration2014}. Unless otherwise specified, distance
units are comoving (and may be explicitly specified as such by the
prefix `c'). Magnitudes are quoted in the AB system \citep{oke1983}.

\section{Model}
\label{sec:mod-igm}

\subsection{The Sherwood-Relics simulations}
\label{sec:sherw-relics-simul}
The Sherwood-Relics simulation suite \citep{puchwein2023} is a set of
high-resolution cosmological hydrodynamical simulations performed with
a modified version of the {\sc p-gadget-3} code (itself a modified
version of {\sc gadget-2}, described in \citealt{springel2005}). Here
we use the $236\rm\,cMpc$ (\ie $160~\cmpch$) box, containing
$2\times 2048^3$ particles, with a dark matter particle mass of
$M_{\rm d} = 5.07\times10^7~\msol$ and a gas particle mass of
$M_{\rm g} = 9.41\times10^6~\msol$. The simulation suite is primarily
focused on studying the intergalactic medium (IGM) and does not
include any subgrid prescription for galaxy formation. It instead
employs a computationally-efficient scheme to remove dense gas,
converting all gas particles with overdensity
$\rho_{\rm g}/\bar{\rho}_{\rm g} = \Delta > 1000$ and temperature
$T<10^5~{\rm K}$ to collisionless star particles (the `quick \lya{}'
approach, \citealt{viel2004}). Halo finding is performed using the
inbuilt friends-of-friends (FoF) halo finder \citep{springel2005}, with the
standard linking length of 0.2 times the mean interparticle
spacing. Note that whenever we discuss `halo mass', we refer to total
FoF group mass
$M_h=M_{h, {\rm d}}+ M_{h, {\rm g}}+M_{h, {\star}}$ where
$M_{h, {\rm d}}$, $M_{h, {\rm g}}$, and $M_{h, {\star}}$ are the dark
matter, gas and stellar mass components of the halo, respectively.

Sherwood-Relics uses a novel hybrid radiative transfer scheme to
capture the hydrodynamical effect of patchy reionisation; we briefly
review the scheme here, but for the full details see
\citet{puchwein2023}. In this method, the monochromatic radiative
transfer of ionising photons is performed using {\sc aton}
\citep{aubert2008}. {\sc aton} is run in post-processing mode (\ie the
evolution of the density field is not coupled to the radiation) on a
base {\sc p-gadget-3} simulation, where the density field is refreshed
every 40~Myr. The spatially-varying \hi{} photoionisation rate is
extracted from this {\sc aton} simulation and used as in input to a
new {\sc p-gadget-3} simulation, thereby allowing the patchy nature of
reionisation to be captured. The luminosity of each source is
proportional to its halo mass, and the minimum mass of haloes that
host ionising sources is $M_h>10^{9.17}~{\rm M}_\odot$ (\ie
$>10^9~\msolh$). The emissivity of each halo is determined by fixing
the redshift evolution of the global emissivity, with the latter
calibrated to match the observed \lya{} forest transmission
\citep{kulkarni2019, keating2020}. The simulation used in this work
completes reionisation at $z=5.3$ and is matched to the
\citet{bosman2022} mean \lya{} transmission constraints (see fig.~1 in
\citealt{conaboy2025}).

Sightlines are drawn through the simulation volume with a spatial
resolution of $115.2~{\rm ckpc}$ to obtain the quantities needed to
produce mock \lya{} absorption spectra. The \lya{} optical depths are
computed using the \citet{tepper-garcia2006} approximation to the
Voigt line profile, including the effects of peculiar velocities.

\subsection{Populating haloes with {\rm \oiii{}} emitters}
\label{sec:populating-haloes}
In order to relate the haloes in the Sherwood-Relics simulation to
galaxies, we construct an empirical relation between halo mass and
observed galaxy properties. We achieve this by using the abundance
matching technique \citep{vale2004, cooray2005} to populate haloes
with \oiii{}-emitting galaxies, an approach that has already been
widely applied to high redshift galaxies (\eg \citealt{bouwens2008,
 lee2009, trenti2010, trac2015}).

We shall first abundance match the halo mass function of our
simulation to the observed UV luminosity function, before using a
scaling relation to obtain the \oiii{} luminosity. We choose to
abundance match to the UV luminosity function and convert to \oiii{}
luminosities -- as opposed to matching directly the \oiii{} luminosity
function -- as the UV luminosity function at $z\sim 6$ has been well
constrained using deep and wide {\em HST} observations
\citep[][hereafter \citetalias{bouwens2021}]{bouwens2021}. This
contrasts with the small survey area and limited sample size of
current \oiii{} emitter surveys carried out with {\em JWST}
(\citealt{matthee2023}, hereafter \citetalias{matthee2023}), although
larger {\em JWST} surveys have characterised the \oiii{} luminosity
function at $z > 7$ \citep{meyer2025}. We also experimented with
abundance matching directly to the \oiii{} luminosity functions
provided by \citetalias{matthee2023} at $z\sim 6$ and \citet{meyer2025}
at $z > 7$, but found that: the lack of turnover at high luminosities
in the \citetalias{matthee2023} Schechter fits produced an
unreasonable mapping; and that matching to the $z=7.1$
\citet{meyer2025} luminosity function produced a mapping consistent
with our two-step method to mostly $<1\sigma$, but the central redshift of
that luminosity function is outside the range we consider here
($5\leq z \leq 7$). We therefore use our two-step matching process
throughout this work.

\subsubsection{Abundance matching}
\label{sec:uv-mag}
We relate UV luminosities to halo masses by assuming that the abundance
of objects measured by each tracer should be equal, \ie
\begin{equation}
  \label{eq:abund-match}
  \int^\infty_{L_{\rm UV}}\phi(L_{\rm UV}', z)\ud L_{\rm UV}' = \epsilon_{\rm DC}(M_h, z) \int^\infty_{M_h} n(M_h',z)\ud M_h',
\end{equation}
where $L_{\rm UV}$ is the UV luminosity, $\phi(L_{\rm UV}, z)$ is the UV
luminosity function, $M_h$ is halo mass, $n(M_h, z)$ is the
differential halo mass function, and
$0 \leq \epsilon_{\rm DC}(M_h, z) \leq 1$ is the duty cycle, which in general can
be mass and redshift dependent. For the UV luminosity function, we use
the Schechter function fit in \citetalias{bouwens2021}, which is
presented in magnitude form as
\begin{equation}
  \label{eq:phi_M_UV}
  \begin{aligned}
    \phi(M_{\rm UV})\,\ud M_{\rm UV} = 0.4 \ln 10 \,\phi^\star\left[10^{-0.4(M_{\rm UV}-M^\star_{\rm UV})}\right]^{\alpha + 1}\times \\
    \exp\left[-10 ^{-0.4(M_{\rm UV}-M^\star_{\rm UV})}\right]\,\ud M_{\rm UV},
  \end{aligned}
\end{equation}
where $\phi^\star$ is the normalisation of the Schechter function,
$M_{\rm UV}^\star$ is the characteristic luminosity, and $\alpha$ is the
faint-end slope. We match to the $z=5.9$ fit, with parameters
\begin{equation}
    \begin{gathered}
      \label{eq:B21}
      \phi^\star = 0.51^{+0.12}_{-0.10} \times 10^{-3}~{\rm cMpc}^{-3}, \\
      M_{\rm UV}^\star =-20.93 \pm 0.09, \\
      \alpha = -1.93 \pm 0.08.
    \end{gathered}
\end{equation}
We convert this to the Schechter function in its luminosity form
\begin{equation}
  \label{eq:phi_L}
  \phi(L_{\rm UV})\, \ud L_{\rm UV} = \frac{\phi^\star}{L_{\rm UV}^\star} \left(\frac{L_{\rm UV}}{L_{\rm UV}^\star}\right)^\alpha \exp \left(-\frac{L_{\rm UV}}{L^\star_{\rm UV}}\right)\, \ud {L_{\rm UV}},
\end{equation}
where $L^\star_{\rm UV}$ is the characteristic luminosity, obtained from
$M^\star_{\rm UV}$ using the definition of AB magnitude
\begin{equation}
  \label{eq:ab}
  M_{\rm UV} = -2.5 \log_{10}\left( \frac{L_{\rm UV,\nu}}{{\rm erg}\,{\rm s}^{-1}\,{\rm Hz}^{-1}}\right) + 51.6.
\end{equation}
Here $L_{{\rm UV},\nu}$ is the monochromatic UV luminosity at
$1600~\angstrom$. Note there is an
analytic solution to the integral on the left-hand side of
\eref{eq:abund-match},
$\int^\infty_{L_{\rm UV}}\phi(L_{\rm UV}', z)\ud L'_{\rm UV} = \phi^\star\Gamma(\alpha+1,L_{\rm
  UV}/L^\star_{\rm UV})$, where $\Gamma$ is the upper incomplete gamma
function, avoiding the need to carry out a numerical
 integration. 
 
 We incorporate the uncertainties on the parameters in \eref{eq:B21}
 by generating 1024 draws of the luminosity function, independently
 sampling the uncertainties on each parameter from a Gaussian (or two
 half-Gaussians when the errors are asymmetric,
 \citealt{barlow2003}). From each of these draws of the luminosity
 function, we generate a mapping \lmh{} from which we can estimate the
 median and $68~{\rm per}~{\rm cent}$ confidence intervals. The
 uncertainties on each parameter are not independent (cf.~fig.~6 in
 \citetalias{bouwens2021}), but we do not have access to the original
 distribution from which the uncertainties in \eref{eq:B21} are
 calculated. To account for this, we follow \citet{trac2015} in
 multiplying the uncertainties on each parameter by $0.7$, such that
 the $1\sigma$ confidence interval on $\phi(M_{\rm UV})$ is approximately
 equal to the $1\sigma$ uncertainty on the binned data points from
 \citetalias{bouwens2021}.

 For the differential halo mass function, $n(M_h,z)$, we use the
 analytic \citet{sheth2001} halo mass function. This agrees with the
 halo mass function measured from our simulation to better than
 $10~{\rm per~cent}$ (and mostly better than $1~{\rm per~cent}$) over
 the redshifts and masses we consider. Using the analytic halo mass
 function to construct the $L_{\rm UV}(M_h)$ mapping -- as opposed to
 the mass function from our simulation -- has the advantage that it
 does not suffer from resolution or finite box size effects at the low
 and high-mass ends, respectively. Note that the analytic mass
 function is only used to construct the $L_{\rm UV}(M_h)$ mapping, and
 all other results in this work use the haloes obtained from the
 Sherwood-Relics simulation.
 
\begin{figure*}
  \centering
  \includegraphics[width=\linewidth]{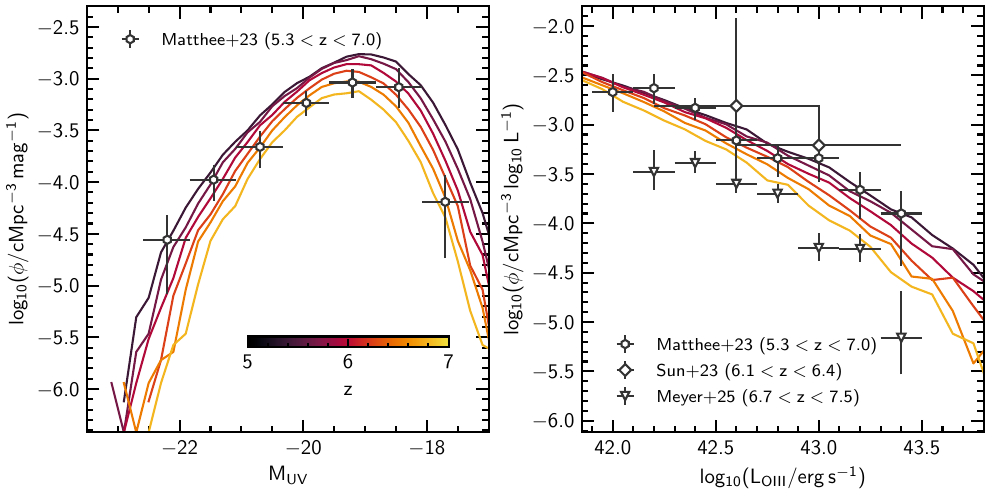}
  \vspace{-0.5cm}
  \caption{Simulated UV luminosity functions (left panel, lines), with
    observational data from \citetalias{matthee2023} (grey hexagons). We
    also show the corresponding \oiiia{} luminosity functions (right
    panel, lines) with observational data from \citetalias{matthee2023}
    (grey hexagons), \citet{sun2023} (grey diamonds), and
    \citet{meyer2025} (grey triangles). The simulated luminosity
    functions are shown every $\Delta z=0.4$ between $z=5$ and $z=7$, where
    the line colour indicates the redshift.}
  \label{fig:uvlf}
\end{figure*}

Lastly, for the duty cycle we use the \citet{trenti2010} `improved
conditional luminosity function', which has been demonstrated to
perform well for abundance matching studies at high redshift
\citep[e.g.][]{weinberger2019, keating2020a, chakraborty2026,
 maitra2025}. The duty cycle is defined for a halo mass $M_h$ and at
a redshift $z$ as
\begin{equation}
  \label{eq:duty-cycle}
  \epsilon_{\rm DC}(M_h, z) = \frac{\displaystyle\int^\infty_{M_h}\left[n(M'_h, z) - n(M_h', z_{\Delta t})\right]\ud M_h'}{\displaystyle\int^\infty_{M_h}n(M_h', z) \ud M_h'},  
\end{equation}
where $z_{\Delta t}$ is the redshift corresponding to the time
$t(z) - \Delta t$, $t(z)$ is the age of the Universe at redshift $z$ and
$\Delta t$ is the timescale over which galaxies are UV
bright. \citet{trenti2010} chose $\Delta t=200~{\rm Myr}$, but in this work
we choose a shorter timescale $\Delta t=50~{\rm Myr}$. Our choice of
$\Delta t$ is motivated by the fact that the escape fraction has been shown
to fluctuate rapidly over timescales $\lesssim100~{\rm Myr}$ in
high-resolution studies of galaxy formation and feedback during the
epoch of reionisation \citep{rosdahl2022}. Similar timescales were
used by \citet{chakraborty2026} who find good agreement with {\em
  JWST} UV luminosity functions and galaxy clustering measurements
when using a timescale that varies with both redshift and halo mass,
with $\Delta t\approx45~{\rm Myr}$ at $z=6$ for
$10^{10}~{\rm M}_\odot$ haloes. We explore the impact of choosing
different duty cycle timescales in \sref{sec:impact-duty-cycle},
finding that the choice of timescale has a negligible effect on our
results.

With a sample of galaxies with abundance matched UV magnitudes in
hand, we next convert the UV magnitudes into \oiii{} luminosities. We
follow \citet{meyer2024} by calculating the ratio $\loiii/L_{\rm UV}$
that maps the abundance-matched \citetalias{bouwens2021} UV luminosity
function to the observed \citetalias{matthee2023} UV luminosity function of
the \oiii-selected galaxies. Assuming a limiting flux sensitivity,
\begin{equation}
  \label{eq:flim} 
  f_{\rm lim} = \frac{L_{\rm lim}}{4\pi D_L^2},
\end{equation} 
where $L_{\rm lim}$ is the limiting \oiiia{} luminosity and $D_L$ the
luminosity distance, we fit for the $\loiii/L_{\rm UV}$ relation that
reproduces the observed UV luminosity function of the \oiii{} emitters
from the abundance-matched \citetalias{bouwens2021} UV luminosity
function. The limiting line sensitivity depends on the survey, with
EIGER reporting a minimum of
$0.6\times10^{-18}~{\rm erg\, s^{-1}\, cm^{-2}}$ for a $3\sigma$ detection
\citepalias{matthee2023}, while ASPIRE report a minimum of
$2.0\times10^{-18}~{\rm erg\, s^{-1}\, cm^{-2}}$ for a $5\sigma$ detection
\citep{wang2023}. As a compromise, we adopt a limiting line
sensitivity of $1.0\times10^{-18}~{\rm erg\, s^{-1}\, cm^{-2}}$ (the same
as used in \citealt{meyer2024}, who use the FRESCO
survey).\footnote{At $z=6$, a limiting flux of
  $1.0\times10^{-18}~{\rm erg\, s^{-1}\, cm^{-2}}$ translates to an
  \oiiia{} luminosity of
  $\loiii > 4.2 \times 10^{41}~{\rm erg\, s^{-1}}$.} We also experimented
with using $f_{\rm lim}=0.6$ and
$2.0~\times 10^{-18}~{\rm erg\, s^{-1}\, cm^{-2}}$, and found that our
results for the galaxy--\lya{} transmission cross-correlation are
insensitive to the choice of $f_{\rm lim}$.

As with \citet{meyer2024}, we assume that the ratio
$\loiii/L_{\rm UV}$ follows a Gaussian distribution with mean
$\mu = \mu_0 + a \left( \log_{10} L_{\rm UV} - \log_{10} L_{\rm
    UV,0}\right)$, where $L_{\rm UV}$ and $L_{\rm UV,0}$ are both in
units of ${\rm L}_\odot$, and a standard deviation $\sigma$. In this model, the
predicted $L_{\rm \oiii}$ will fall below $L_{\rm lim}$ (defined in
\eref{eq:flim}) for some fraction of galaxies, rendering those
galaxies undetected. This fraction of undetected galaxies should
increase with $M_{\rm UV}$ to match the fall-off in the UV luminosity
function of \oiii-selected galaxies at $M_{\rm UV}\gtrsim -20$. We set
$\log_{10}L_{\rm UV, 0}=10.93$, as this is the luminosity above which
the \citetalias{matthee2023} UV luminosity function data are in best
agreement with the \citetalias{bouwens2021} UV luminosity function. Using
our $z=6$ abundance-matched UV luminosity function, we then fit for
$\mu_0$, $a$, and $\sigma$ using the \citetalias{matthee2023} data to obtain the
following mean $\loiii/L_{\rm UV}$ ratio
\begin{equation}
\label{eq:O3UVr}
\begin{aligned} \log_{10}\left(\frac{\loiii}{L_{\rm
UV}}\right)& = \\ -1.42 & + 0.49 \left( \log_{10} L_{\rm UV} -
10.93\right) \pm 0.36.
\end{aligned}
\end{equation} 
The parameter $\sigma$ controls the range of $L_{\rm \oiii}$ a given
$M_{\rm UV}$ can take. The $68~{\rm per~cent}$ confidence interval
(holding the other parameters fixed at their best-fit values) of the
fitted $\sigma=0.36^{+0.07}_{-0.04}$, and varying $\sigma$ within this range
means that we are consistent with the $L_{\rm \oiii}$--$M_{\rm UV}$
relation from \citetalias{matthee2023} for $M_{\rm UV}\lesssim -19$, but that
we predict lower \oiii{} luminosities for fainter magnitudes.

Using \eref{eq:O3UVr} we then generate \oiiia{} luminosities from the
UV magnitudes of the abundance matched galaxies, removing any galaxies
from the catalogue with a flux that falls below the limiting line
sensitivity of $1.0\times10^{-18}~{\rm erg\, s^{-1}\, cm^{-2}}$. After
applying all of the above steps, we obtain a catalogue of \oiii{}
emitters derived from the dark matter haloes in our simulation.

To summarise, in this work we use or compare to three types of
luminosity functions:
  \begin{itemize}
  \item we first abundance match to the UV luminosity function of
    galaxies selected by their UV luminosity \citepalias{bouwens2021};
  \item we then fit the UV luminosity function of galaxies selected by
    their \oiiia{} emission \citepalias{matthee2023} using
    Eq.~(\ref{eq:O3UVr});
  \item lastly, we predict the \oiiia{} luminosity function of galaxies
    selected by their \oiiia{} emission
    (\citetalias{matthee2023}, \citealt{sun2023}, and
    \citealt{meyer2025}).
  \end{itemize}

\subsubsection{Galaxy abundance}
We now present some consistency checks of our simulated galaxy
catalogue. In \fref{fig:uvlf} we show the UV and \oiiia{} luminosity
functions for our full, flux-limited catalogue, \ie making no cuts to
mimic survey volume, between $5 \leq z \leq 7$. In the left panel we show
the UV luminosity function compared to the observed UV luminosity
function at $5.33 < z < 6.96$ from \citetalias{matthee2023}, finding
excellent agreement with the observed data {\em by construction},
particularly around $z\approx 6$. The turnover in the UV luminosity function
for $M_{\rm UV} \gtrsim -20$ is due to incompleteness at the faint end of
the observed \oiiia{} luminosity function and the scatter in the
$L_{\rm \oiii}/L_{\rm UV}$ ratio, which can mean that the \oiii{} line
for a given $L_{\rm UV}$ can fall below the limiting flux sensitivity,
rendering the galaxy undetected. It is also worth noting here that,
although we carry out the $L_{\rm UV}(M_h)$ mapping at fixed redshift
($z=5.9$), the simulated luminosity functions evolve with
redshift. This is due to the evolution of the underlying halo mass
function in the Sherwood-Relics simulation, the redshift dependence of
the duty cycle (see \eref{eq:duty-cycle}), and the redshift evolution
of $L_{\rm lim}$ at fixed $f_{\rm lim}$ (see \eref{eq:flim}).

In the right panel of \fref{fig:uvlf}, we show the \oiiia{} luminosity
function with observational estimates from \citetalias{matthee2023},
\citet{sun2023} ($6.11 < z< 6.35$), and \citet{meyer2025}
($6.75 < z < 7.50$). We find good agreement with \citetalias{matthee2023},
particularly at $z \lesssim 6.4$, and note that this is no longer by
construction; the \oiii{} luminosity function from \citetalias{matthee2023}
never features in our fitting. Instead, this good agreement
demonstrates the validity of the $\loiii/L_{\rm UV}$ ratio modelling
discussed in \sref{sec:uv-mag}, as well as the consistency
 between the UV and \oiii{} luminosity functions reported by
 \citetalias{matthee2023}. We predict slightly lower abundances than
\citet{sun2023} (although still within $1\sigma$ for
$z \lesssim 6.8$). However, we note that the \citet{sun2023} sample comprises
only four \oiii{} emitters. We overshoot the \citet{meyer2025}
measurement at all redshifts and luminosities except for the point at
$\log_{10}(\loiii/{\rm erg\, s^{-1}}) = 43.2$ and $z=7$. This is
perhaps not too surprising given that the mean redshift of the
\citet{meyer2025} dataset is $\langle z \rangle = 7.1$, \ie higher than any
redshift we show, and that the abundance of \oiii{} emitters at all
luminosities tends to decrease with increasing redshift.

\subsubsection{Galaxy clustering}
\label{sec:galaxy-clustering}
Two {\em JWST} programmes, EIGER and ASPIRE, have recently reported
measurements of the clustering of high-redshift, \oiii{}-selected
galaxies \citep{eilers2024, huang2026}. They quantify the strength of the
clustering using the volume-averaged projected autocorrelation function
\citep{hennawi2006a, garcia-vergara2017, eilers2024},
\begin{equation}
  \begin{aligned}
  \label{eq:chi-V}
    \chi_V(r_{\perp, {\rm min}}, r_{\perp, {\rm max}})& = \\
    \frac{2}{V} \int^{r_{\perp, {\rm max}}}_{r_{\perp, {\rm min}}} &\int^{Z_{\rm max}}_0 \xi(r_\perp, Z) 2 \pi r_\perp \ud r_\perp \ud Z,
  \end{aligned}
\end{equation}
where $r_\perp$ is the radial distance perpendicular to the line-of-sight,
$Z$ is the line-of-sight separation, and $\xi(r_\perp, Z)$ is the
correlation function binned in both the radial and line-of-sight
directions. The volume of each cylindrical shell, $V$, is given by
\begin{equation}
  V=2 Z_{\rm max} \pi (r_{\perp,{\rm max}}^2 - r_{\perp,{\rm min}}^2),
\end{equation}
where $Z_{\rm max}$ is the maximum line-of-sight
separation. 

\citet{eilers2024} determine $Z_{\rm max}$ through fixing the maximum velocity difference by which pairs of
 galaxies can be separated $v_{\rm max}=|\Delta v|$, giving
\begin{equation}
  \label{eq:Zmax}
  Z_{\rm max} = \frac{v_{\rm max}(1+z)}{H(z)},
\end{equation}
where $H(z)$ is the Hubble parameter at $z$, and
$v_{\rm max}=1000~{\rm km ~s}^{-1}$, corresponding to
$Z_{\rm max}\approx10.3~{\rm cMpc}$ (\ie $7~\cmpch$) at $z=6$. Instead, \citet{huang2026} fix
$Z_{\rm max}\approx 10.3~{\rm cMpc}$, thus implicitly allowing
$v_{\rm max}$ to vary with redshift. We follow \citet{eilers2024}, and
fix $v_{\rm max}=1000~{\rm km\, s^{-1}}$, but note that the difference
should be small (\eg \citealt{huang2026} obtain
$v_{\rm max}=1037~{\rm km\, s^{-1}}$ at $z=6.5$). When computing
autocorrelation functions from the EIGER results, \citet{eilers2024}
include all galaxies with $\log_{10}(\text{\loiii}/\ergs) \geq 42.0$, and
\citet{huang2026} use a lower limit of
$\log_{10}(\text{\loiii}/\ergs) > 41.8$. However, for the EIGER
survey, the completeness at this luminosity is only
$\sim40~{\rm per~cent}$, rising to $\sim 80~{\rm per~cent}$ for
$\log_{10}(\text{\loiii}/\ergs) \geq 42.4$ \citep{kashino2026}. For this
reason, we follow \citet{pizzati2024} who only include galaxies with
$\log_{10}(\text{\loiii}/\ergs) \geq 42.4$ when comparing their
clustering model to the EIGER observations. Making this
 minimum cut in luminosity yields a mean luminosity
 $\langle \log_{10}(\loiii{}/{\rm erg\, s^{-1}})\rangle =42.73$ at
 $z=6$, which is 0.20~dex larger than the mean luminosity of 42.53
 reported by \citet{huang2026}, while a cut of
 $\log_{10}(\text{\loiii}/\ergs) > 41.8$ would yield a mean
 luminosity of 42.20, \ie 0.33~dex smaller than
 \citealt{huang2026}). 
\begin{figure}
  \centering
  \includegraphics[width=\columnwidth]{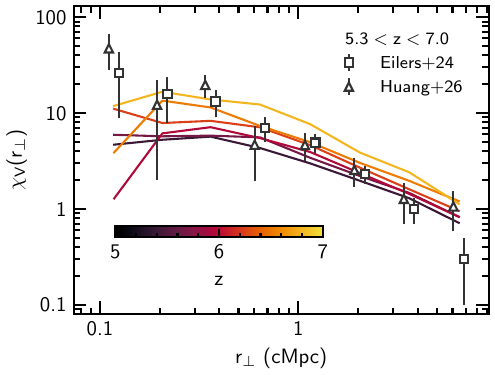}
  \vspace{-0.5cm}
  \caption{Simulated \oiiia{} volume-averaged projected autocorrelation
    functions (solid) for all haloes in the catalogue, compared to
    observational results from \citet{eilers2024} (grey squares) and
    \citet{huang2026} (grey triangles), both covering the redshift
    range $5.3< z < 7.0$ and using the same binning (although for
    clarity, we shift the \citet{huang2026} points slightly left and
    the \citet{eilers2024} points slightly right). We include all
    galaxies with $\log_{10}(\text{\loiii}/{\rm erg~s}^{-1}) >
    42.4$. The line colours are the same as in \fref{fig:uvlf}.}
  \label{fig:chiv}
  
\end{figure}
To estimate autocorrelation functions from our catalogues, we use {\sc
  corrfunc} \citep{sinha2020} with the Landy--Szalay estimator
\citep{landy1993}.

In \fref{fig:chiv} we show the volume-averaged projected
autocorrelation function at $5 \leq z \leq 7$ for all galaxies with
$\log_{10}(\text{\loiii}/\ergs) \geq 42.4$ in our catalogue, again making
no spatial cuts to mimic the EIGER or ASPIRE surveys, along with the
observational data from \citet{eilers2024} (calculated for all
galaxies with $\log_{10}(\text{\loiii}/\ergs) \geq 42.0$) and
\citet{huang2026} (calculated for all galaxies with
$\log_{10}(\text{\loiii}/\ergs) > 41.8$). We find that the strength of
clustering in our model generally increases across all spatial scales
as redshift increases. For $z\lesssim 6.2$, we tend to find weaker clustering
than \citet{eilers2024} for $r_\perp \lesssim 0.4~{\rm cMpc}$. We predict a
flatter slope of $\chi_V$ over $r_\perp$ than \citet{eilers2024}. We find
better agreement with the clustering measurements of \citet{huang2026}
over a range of redshifts, except at $r_\perp\approx 0.3~{\rm cMpc}$ where only
our $z=7$ estimate is in agreement with the data, and in the smallest
radial bin at $r_\perp \approx 0.1~{\rm cMpc}$ where none our estimates are in
agreement.
 
Most noticeably, our models predict that the slope of $\chi_V$ should
flatten, and even turn over, for $r_\perp \lesssim 0.4~{\rm cMpc}$. This is
contrary to the observational estimates, which predict that $\chi_V$
should continue to rise for these small $r_\perp$. This is largely a
consequence of the technique that we use to identify haloes in our
simulation, FoF, which identifies haloes by linking groups of nearby
particles, and so does not allow for the possibility that a dark
matter halo could host more than one galaxy. This is a well-known
problem when modelling the correlation function of galaxies by
populating dark matter haloes, where the phrase `one-halo term' is
commonly used to to describe the enhanced contribution on small scales
from multiple galaxies in the same dark matter halo
\citep[\eg][]{cooray2002}. We have verified that the missing
clustering amplitude on small scales is indeed caused by the missing
one-halo term by reanalysing one of our snapshots with {\sc rockstar}
\citep{behroozi2013}, a halo finder that provides information about
substructure. The $\chi_V$ calculated from this catalogue (see
\sref{sec:one-halo-term}) is in good agreement with the data for
$r_\perp \lesssim 0.4~{\rm cMpc}$. Note we do not use the {\sc rockstar}
catalogue for our main analysis, because doing so would break the
correspondence between halo and ionising source location, since the
FoF catalogue was used to determine the locations of ionising sources
in the radiative transfer simulation.

\section{Galaxy--Lyman $\alpha$ transmission cross-correlation}
\label{sec:galaxy-lyman-alpha}
The reasonable agreement between our simulated galaxy catalogue and
the observed \oiii{} luminosity function and autocorrelation function
enables us to discuss the impact of galaxy selection on the
galaxy--\lya{} transmission cross-correlation. In addition to the
source modelling discussed in \sref{sec:populating-haloes}, we now
additionally mimic the characteristics of the {\em JWST} surveys,
namely, the galaxy sample (\ie number and luminosity) and the survey
geometry (\ie field-of-view and sightline length). When designing our
mock survey for exploring the galaxy--\lya{} forest transmission
cross-correlation, we focus on the latest results from the ASPIRE
survey \citepalias[see Tab.~\ref{tab:surveys} for details on the
configuration used]{kakiichi2025}.

\begin{table}
  {
    \centering
    \caption{Survey configurations used in
      \sref{sec:galaxy-lyman-alpha} \citepalias{kakiichi2025} and
      \sref{sec:galaxy-density-lyman} \citepalias{zhu2026}.}
    \label{tab:surveys}

    \begin{tabular}{ l c c c }
      \hline
      Ref & $N_{\rm QSO}^*$ & FoV$^\dagger$ (cMpc) & \lya{} forest extent$^\ddagger$ (cMpc) \\
      \hline
      \citetalias{kakiichi2025} & 5 & 5.3--5.5, 12.3--12.8 & 423--466 \\
      \citetalias{zhu2026} & 2 & 24.7, 21.1 & 74 \\ 
      \hline
    \end{tabular}}
  $^*$ Number of sightlines used per realisation\\
  $^\dagger$ Field-of-view including detector gaps. For \citetalias{kakiichi2025} we list the range from $z=5.6$ to $z=6.4$, while for \citetalias{zhu2026} we list the FoV at $z=5.6$.\\
  $^\ddagger$ For \citetalias{kakiichi2025} we list the range across all five sightlines, while the extent for \citetalias{zhu2026} corresponds to $50~\cmpch$.
\end{table}

\subsection{Galaxy sample}
\label{sec:galaxy-sample-K25}
\begin{figure}
  \centering
  \includegraphics[width=\columnwidth]{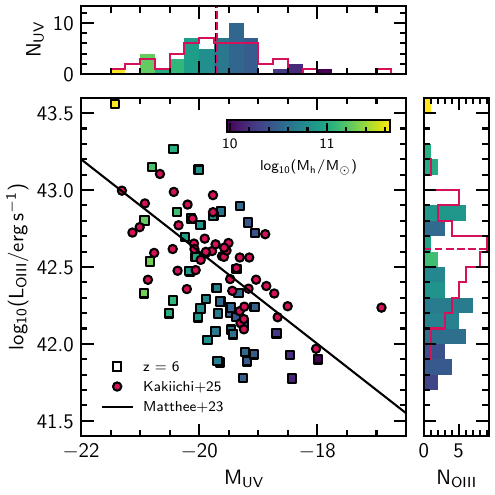}
  \caption{Relation between UV magnitude and \oiiiab{} luminosity for
    a single mock realisation of an ASPIRE-like survey at $z=6$
    (coloured squares), where the colour of the square shows
    ${\rm log}_{10}(M_h/{\rm M}_\odot)$. Also shown are the observations
    from ASPIRE (red circles, \citetalias{kakiichi2025}) and the relation
    from \citetalias{matthee2023}. Side panels show the marginal
    distributions for ASPIRE (red line) and this work (coloured filled
    bars), where the colour of the filled bars shows
    $\langle \log_{10}(M_h/{\rm M}_\odot) \rangle$ for that bin. The dashed lines
    indicate the means.}
  \label{fig:M_UV-L_OIII}
\end{figure}

The distribution of UV magnitudes probed by the ASPIRE survey (see the
red histogram on the top panel of \fref{fig:M_UV-L_OIII}) is well-fit
by a Gaussian with mean $\langle M_{\rm UV} \rangle =-19.7$ and standard deviation
$\sigma_{M_{\rm UV}}=0.7$. For each of the five quasar sightlines in the
ASPIRE survey we therefore draw \oiii{} emitters from this Gaussian
such that the mean number of \oiii{} emitters in a typical ASPIRE-like
realisation $\langle N_{\rm \oiii} \rangle$ is similar to that in
\citetalias{kakiichi2025}. At $z = 6.4$, $6.0$, and $5.6$ we find
$\langle N_{\rm \oiii}\rangle=7.2$, 8.8, and 10.2, compared to
$\langle N_{\rm \oiii}\rangle =9.8$ in \citetalias{kakiichi2025}, although our
results are not very sensitive to this choice. We have verified that
performing the calculation with $\langle N_{\rm \oiii} \rangle=17.4$ at
$z=6$ does not change our results for the galaxy--\lya{} transmission
cross-correlation, consistent with the findings of
\citet{garaldi2025a} (see their fig.~5). We furthermore found that
fixing the number of \oiii{} emitters per sightline to the ASPIRE
values breaks the correspondence between the number density of
galaxies and the galaxy--\lya{} transmission cross-correlation. By
approximately matching only $\langle N_{\rm \oiii}\rangle$, we allow for this
natural variation in the number density of galaxies. The reason that
we have to sample galaxies at all, as opposed to using all galaxies in
an ASPIRE-like volume, is that the \citetalias{matthee2023} \oiii{}
luminosity function predicts a larger number of galaxies compared to
the ASPIRE fields by a factor $\sim 4$ (see also \citetalias{zhu2026}).
An example of the resultant $\loiii$--$M_{\rm UV}$ relation is shown in
\fref{fig:M_UV-L_OIII}, along with the observed relation from
\citetalias{kakiichi2025} and the linear fit from
\citetalias{matthee2023}. Note that in \fref{fig:M_UV-L_OIII} we show
the \oiiiab{} luminosity, as opposed to the \oiiia{} luminosity used
throughout the rest of this work, and relate the two using the
theoretical \oiiib{}:\oiiia{} line ratio of 1:2.98 \citep{storey2000}.

By design, our model has a similar distribution in $M_{\rm UV}$ to
\citetalias{kakiichi2025}, however the galaxies tend to have slightly lower
$\loiii$ than both \citetalias{matthee2023} and \citetalias{kakiichi2025} due to
the steeper $\loiii$--$M_{\rm UV}$ relation required to reproduce the
\oiii{}-selected \citetalias{matthee2023} UV luminosity function from the
(abundance-matched) \citetalias{bouwens2021} UV luminosity function. This
leads to a mean luminosity
$\langle\log_{10} (L_{\rm \oiii}/{\rm erg\,s^{-1}})\rangle\approx 42.5$ in our model,
compared to $\approx 42.6$ in \citetalias{kakiichi2025}. As a consequence of the
abundance matching process, the mean halo mass
$\langle \log_{10} (M_h/{\rm M}_\odot)\rangle$ (indicated by the colour of the points
and filled bars in \fref{fig:M_UV-L_OIII}) monotonically increases
with $M_{\rm UV}$, but the scatter in the $\loiii/L_{\rm UV}$ ratio
means that $\langle \log_{10} (M_h/{\rm M}_\odot)\rangle$ does not increase strictly
monotonically with $\loiii$. However, the general trend is that
galaxies brighter in $\loiii$ occupy more massive haloes. We note that
the shape of the $\loiii$--$M_{\rm UV}$ relation is sensitive to both
the choice of $f_{\rm lim}$ and the redshift at which the fitting of
\eref{eq:O3UVr} is carried out. Increasing $f_{\rm lim}$ tends to
increase the amplitude of the $\loiii$--$M_{\rm UV}$ relation, meaning
that a galaxy of a given $M_{\rm UV}$ will tend to be brighter in
$\loiii$. Holding $f_{\rm lim}$ fixed, but increasing the redshift at
which the fit is carried out tends to result in a shallower
$\loiii$--$M_{\rm UV}$ relation.

\subsection{Survey geometry}
\label{sec:survey-geometry-K25}
The ASPIRE survey employs single NIRCam pointings, where each of the
modules has an angular size of $2.2'$, with a $44''$ gap between
modules. The quasar sightline is placed in Module A at an offset
relative to the centre of the entire field of view (in both horizontal
and vertical) of $X_{\rm offset}=-60''.5$ and $Y_{\rm offset}=7''.5$
\citep{wang2023}. We also tested a configuration where the quasar
sightline is placed at the centre of the field-of-view, rather than
off-centre, and found that the results for the galaxy--\lya{}
transmission cross-correlation did not change. This survey design is
much simpler than, for example, the multi-visit strategy employed by
EIGER, which leads to a complex completeness map \citep[cf. fig.~4
in][]{kashino2026}. We also note that the efficiency of NIRCam's
Module B grism is $\sim 30{~\rm per cent}$ of that of Module A
\citep{greene2016}, but defer a detailed study of the impact of this --
if any -- to future work.

The extent of the \lya{} forest in each quasar sightline used by
ASPIRE is $\sim450~{\rm cMpc}$, which is almost double our box size of
$236~{\rm cMpc}$, and so we use the periodicity of the simulation
volume to extend our mock data to the required path length.

\subsection{Mock survey}
\label{sec:mock-survey-K25}
We fold all of these observational aspects together to generate a
realisation of a mock survey. We generate five quasar sightlines as in
the ASPIRE programme, where each sightline used in
\citetalias{kakiichi2025} has the same geometric properties -- sightline
length and field-of-view -- in our model. Given the randomness
involved in generating each sightline -- \ie which sightline is chosen
and which haloes from the parent catalogue are selected -- we generate
an ensemble of 1024 realisations of these five sightlines. Each
realisation we call a `survey', and from the ensemble of surveys we
may obtain the mean and distribution of the cross-correlation as a
function of distance from galaxies.

\subsection{Results}
\label{sec:results-K25}
\begin{figure}
  \centering
  \includegraphics[width=\columnwidth]{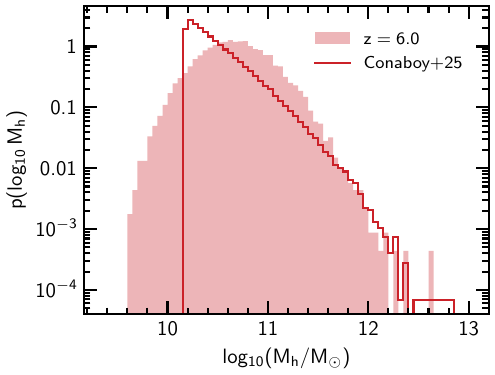}
  \caption{The probability distribution of halo masses used in the
    calculation of the \oiii{} emitter-\lya{} cross-correlation at
    $z=6$ (see \fref{fig:delta_F}). Filled bars indicate the
    distribution used in this work, while the red curve indicates the
    distribution used in \citet{conaboy2025}, \ie all haloes with
    $M_h \ge 10^{10} ~ \msolh$.}
  \label{fig:lMh-dist}
\end{figure}

In \fref{fig:lMh-dist} we first show the distribution of halo masses
used to calculate the galaxy--\lya{} transmission correlation at $z=6$
in \citet{conaboy2025} and of all the 1024 mock ASPIRE-like surveys
described in \sref{sec:mock-survey-K25}. We obtain a mean halo mass of
$\langle\log_{10}(M_h/{\rm M}_\odot)\rangle \approx 10.7$ at
$z=6$, compared to $\approx10.5$ for the sample in \citet{conaboy2025}. We
find that our sample covers roughly three decades in halo mass and
extends down to $\log_{10}(M_h/{\rm M}_\odot)\approx 9.6$, well below the lower
limit of $\log_{10}(M_h/{\rm M}_\odot)\approx10.2$ applied in
\citet{conaboy2025}. The larger scatter in halo masses is attributable
to the scatter intrinsic to the \lmh{} abundance matching (from the
uncertainties in the Schechter fits), as well as the scatter in
$M_{\rm UV}$ sampled by ASPIRE.

\begin{figure}
  \centering
  \includegraphics[width=\columnwidth]{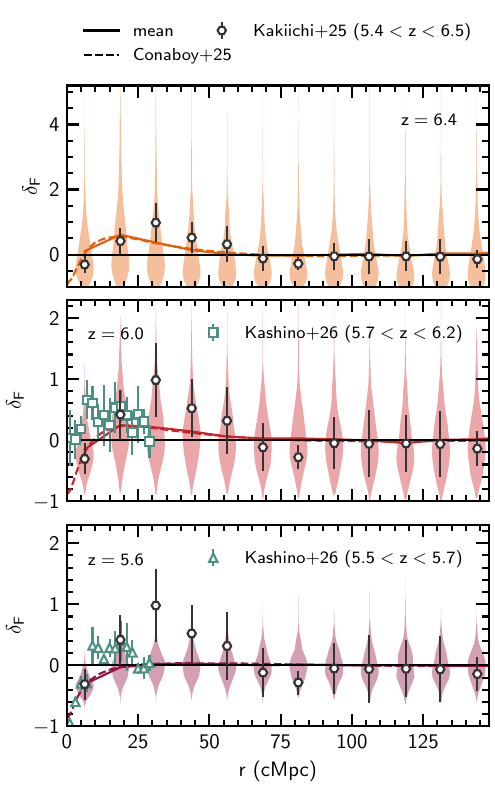}
  \caption{The galaxy-\lya{} transmission cross correlation at
    $z=6.4$, $6.0$ and $5.6$ (top, middle and bottom panels) for our
    full forward model. Solid curves indicate the mean, and the violin
    plots indicate the distribution of $\delta_F$ within each $r$ bin from
    1024 ASPIRE-like realisations using kernel density
    estimation. Note that at $z=6.4$, some of the bins extend to
    larger values of $\delta_F$ than shown here but, to preserve dynamic
    range, we clip the long tail of these distributions and there are
    different $\delta_F$ ranges displayed on the top and bottom two
    panels. Also shown is the modelling from \citet{conaboy2025}
    (calculated for all haloes with $M_h \geq 10^{10}~\msolh$, dotted
    curves) along with observational results from ASPIRE (at
    $5.4 < z < 6.5$, grey hexagons, \citetalias{kakiichi2025}) and
    EIGER (green squares for $5.7 < z < 6.15$ and
    green triangles for $5.5 < z < 5.7$,
    \citealt{kashino2026}).}\label{fig:delta_F}
\end{figure}

In \fref{fig:delta_F} we show the cross correlation in its usual form
\begin{equation}
  \label{eq:delta_F}
  \delta_F = \frac{\langle F(r) \rangle}{\bar{F}} - 1,
\end{equation}
where $\langle F(r) \rangle$ is the average transmission measured at a distance
$r$ from galaxies, and $\bar{F}$ is the global mean transmission
(calculated from all the sightlines available for each snapshot), for
a range of redshifts bracketing those probed by the ASPIRE survey. On
the same figure, we show the results from \citet{conaboy2025} -- which
includes all haloes with $M_h \geq 10^{10}~\msolh$ -- along with
observational results from the ASPIRE \citepalias{kakiichi2025} and EIGER
surveys (\citealt{kashino2026}, an update to the
\citealt{kashino2023} result examined in \citealt{conaboy2025}).

The most striking result of our source and survey modelling in
\fref{fig:delta_F} is the very large scatter in $\delta_F$, particularly at
higher $z$. In addition, the mean of all the realisations at all
redshifts are virtually indistinguishable from \citet{conaboy2025}.
One reason for this large scatter is that \lya{}
transmission spikes are very rare at $z=6.4$, and so the mean
transmission ${\bar F}$ is set by a few very transmissive
pixels. These pixels are well-sampled by \citet{conaboy2025}, where
all sightlines in the simulation box are used. In this work, however,
a subsample of five sightlines are averaged to compute $\delta_F$ -- a
process which is repeated 1024 times to produce the distributions in
\fref{fig:delta_F}. In this setup, the typical (\ie median) mean
transmission computed from the subsampled sightlines in a single
realisation is $\sim 90$~per~cent of the mean transmission in the entire
box at $z=6.4$. By $z=6.0$ and $5.6$, the distribution in transmission
is much more symmetric, and the mean transmission of a typical
realisation is equal to the mean transmission of the entire box. This
effect can be seen in \fref{fig:delta_F}, where the distributions in
$\delta_F$ become more symmetric below $z=6.4$.

Across all redshifts, the large scatter in $\delta_F$ means that all of our
mock surveys are consistent with $\delta_F = 0$ at almost all $r$ (except
marginally for $r\lesssim 10~{\rm cMpc}$ at $z=5.6$). This large scatter
between realisations is consistent with the findings of \citet[see
their figs.~2 and 3]{garaldi2025a}. The variance is important to
consider when interpreting the results of cross-correlation
measurements using small numbers of fields (see fig.~6 in
\citealt{jin2026} for a comparison of single-field cross-correlation
measurements from that work and \citetalias{kakiichi2025}, along with
the two-field measurement from \citetalias{zhu2026}). The scatter
means it is possible to match the large amplitude of the signal in
\citetalias{kakiichi2025} at $z=6.4$ (although our models are also
consistent with $\delta_F < 0$), but it is not possible to shift the
location of the peak from $r \approx 20~{\rm cMpc}$ in our models to
$r\approx 30~{\rm cMpc}$ in the observations. This is because the \oiii{}
emitters in our model tend to live in only slightly more massive
haloes than the mass cut used in \citet{conaboy2025} -- the average
halo mass in this work is $\sim 0.2~{\rm dex}$ higher than in
\citet{conaboy2025}, whereas an increase of $1$--$2~{\rm dex}$ in mass
would be needed to significantly shift the location of peak
transmission (cf. fig.~9 in \citealt{conaboy2025}). We also tend to
predict a smaller value of $\delta_F$ than the \citet{kashino2026} data,
although this difference could be alleviated by assuming a slightly
different evolution of the global neutral fraction (\ie effectively
shifting the \citet{kashino2026} data points to higher redshifts, see
fig. 10 in \citealt{conaboy2025}). Finally, we note that our updated
modelling retains good agreement with the two-point cross-correlation
function of \lya{} emitters (LAEs) and Lyman break galaxies with
\lya{} transmission spikes found by \citet{meyer2020} (cf.\@ fig.~B1
in \citealt{conaboy2025}.)

We conclude that selection cuts based on the detailed modelling of the
\oiii{}-emitter catalogue make little impact on the shape of cross
correlation with \lya{} transmission, but they are critical for
accurately assessing the variance in the measurements. This is
consistent with \citet{garaldi2025a} and their analysis of the
\thesan{} simulations, where these authors find a similar shape for
$\delta_F$ whether they make a cut on \oiii{} flux or other galaxy
properties (see their fig.~9). As discussed in \citet{conaboy2025},
this strongly suggests that the relationship between halo mass,
ionising photon production and the volume averaged neutral fraction in
the intergalactic medium will be primarily responsible for setting the
shape of $\delta_F$, and that simulation volumes
$>160^{3}~h^{-3}\rm\,cMpc^{3}$ are required to better capture the high
mass end of the halo mass function \citep[see also][]{spina2026,
  sarkar2026}.

\section{Relation between galaxy surface density and Lyman $\alpha$ effective optical depth}
\label{sec:galaxy-density-lyman}
We now apply our \oiii{}-emitter model to studying the connection
between galaxy surface density, $\Sigma_{\rm gal}$, and the \lya{} forest
effective optical depth measured in $50~h^{-1}\rm\,cMpc$ windows,
$\tau_{\rm eff}^{50}$. Typically, this measurement has been performed
using \lya{} emitters selected in narrowband surveys
\citep{becker2018,kashino2020,christenson2021,ishimoto2022,christenson2023}. However,
the increasing opacity of the IGM to \lya{} at high redshift makes
identifying \lya{} emission lines difficult. Here we focus on the
recent study by \citetalias{zhu2026}, who perform a {\em JWST} survey
of \oiii{} emitters in the fields of two high-redshift quasars (again,
see Tab.~\ref{tab:surveys} for details on the configuration used in
this section, which matches that used by \citetalias{zhu2026}).

\subsection{Galaxy sample}
\label{sec:galaxy-sample-Z25}
Tiled single NIRCam pointings are employed by \citetalias{zhu2026} with
similar exposure times and \oiii{}-emitter detection methodologies as
in ASPIRE. We therefore assume the same distribution in $M_{\rm UV}$
used in \sref{sec:galaxy-sample-K25}. We also apply the \oiiia{}
luminosity threshold employed by \citetalias{zhu2026}, which only includes
galaxies with $\log_{10}(\loiii/{\rm erg~s^{-1}}) > 42.1$. Finally, we
use the same technique for determining the number of \oiii{} emitters
per sightline as in \sref{sec:galaxy-sample-K25}. We find that this
yields a typical $\langle N_{\rm \oiii} \rangle = 28.0$, compared to
$\langle N_{\rm \oiii}\rangle=29.0$ in \citetalias{zhu2026}.
 
\subsection{Survey geometry}
\label{sec:survey-geometry-Z25}
The NIRCam pointings were carried out in a 4$\times$2 mosaic (see fig.~1 of
\citetalias{zhu2026}), with the quasars placed at the centre. As in
\sref{sec:survey-geometry-K25}, we ignore any effects due to the
differences in throughput between Module A and Module B of NIRCam. The
depth of each NIRCam pointing, corresponding to the redshift interval
$5.3 < z < 6.1$, is $\sim 370{~\rm cMpc}$. As before, we periodically
extend our simulated quasar sightlines to the required length and
randomly periodically shift galaxies along the sightlines.

\subsection{Mock survey}
\label{sec:mock-survey-Z25}
Combining each of these aspects, we generate $2500$ realisations of
each quasar sightline. Given that the length of each quasar sightline contains
multiple $50~\cmpch$ regions (the window used to calculate the
effective optical depth \tauf), we calculate \tauf{} and
$\Sigma_{\rm \oiii}$ over four non-overlapping regions within each
sightline.

\subsection{Results}
\label{sec:results-Z25}
\begin{figure}
  \centering
  \includegraphics[width=\columnwidth]{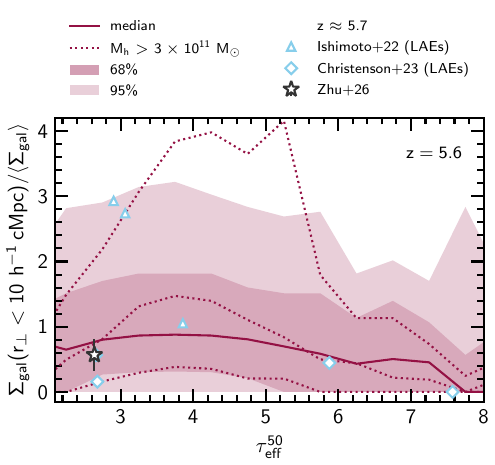}
  \caption{The relation between galaxy surface density and \lya{}
    forest effective optical depth at $z=5.6$. The galaxy surface
    density is measured in a $28~\cmpch$ window, and the \lya{} forest
    effective optical depth in a $50~\cmpch$ window. The median
    relation for the mock \oiii{} sample is shown by the solid curve,
    along with the 68~per~cent (dark band) and 95~per~cent (light
    band) ranges. The dotted curves show the median and 68~per~cent
    range for the case when all haloes with $M_h>3\times10^{11}~\msol$ are
    selected. This mass threshold is chosen such that
    $\langle \Sigma_{\rm gal} \rangle \approx 0.02~(\cmpch)^{-2}$, matching
    \citet{christenson2023}. Data points show the measurements using
    \oiii{} emitters from \citetalias{zhu2026} (black star) and \lya{}
    emitters from \citet{ishimoto2022} (blue triangles) and
    \citet{christenson2023} (blue diamonds), all at $z\approx 5.7$. Note
    that the \citetalias{zhu2026} sample comprises two of the sightlines
    used in the \citet{christenson2023}, one of which is directly
    underneath the \citetalias{zhu2026} data point.}
 \label{fig:Sigma-OIII}
\end{figure}

In \fref{fig:Sigma-OIII} we show the relation between \oiii-emitter
surface density and \tauf{} for our mock survey at $z = 5.6$ (solid
curve with 68 and 95 per cent scatter given by the shading), along
with the observational constraint due to \citetalias{zhu2026}, as well
as constraints from LAEs due to \citet{ishimoto2022}\footnote{Note
  that we use the recalculated values of \tauf{} due to
  \citet{christenson2023}.} and \citet{christenson2023}. Note that
\citetalias{zhu2026} have analysed two of the sightlines studied in
\citet{christenson2023}, and their result (shown as the grey star in
\fref{fig:Sigma-OIII}) sits on top of one of the
\citet{christenson2023} sightlines. Here we have followed
\citetalias{zhu2026} by selecting galaxies in $28~\cmpch$-thick slabs
perpendicular to the line-of-sight at the centre of the $50~\cmpch$
regions used to calculate \tauf{}, and the surface densities
\Sigmaoiii{} and \Sigmaoiiim{} account for the gaps between NIRCam
modules when calculating the survey area. For comparison, we also show
the surface density--\tauf{} relation assuming that every halo with
$M_h > 3\times10^{11}~\msol$ hosts an \oiii{} emitter, which also does not
account for the \citetalias{zhu2026} survey geometry (dotted
curves). This mass limit is chosen such that, typically,
$\langle\Sigma_{\rm gal}\rangle =0.02~(\cmpch)^{-2}$, corresponding roughly to the
typical surface density of galaxies in \citet{christenson2023}. This
facilitates a fairer comparison between our model and the LAE surface
densities from \citet{ishimoto2022} and \citet{christenson2023}. In
\citetalias{zhu2026} the typical surface density of \oiii{} emitters
is instead $\langle \Sigma_{\rm gal} \rangle = 0.015$--$0.016~(\cmpch)^{-2}$.

Our median estimate of the galaxy density--\tauf{} relation is in good
agreement with the observational estimate from \citetalias{zhu2026},
although -- as for \fref{fig:delta_F} -- there is a large scatter. Our
model is able to reproduce the mean density \citet{ishimoto2022} point
at $\tauf \approx 4$, but the extreme galaxy overdensities at
$\tauf \approx 3$ are at the upper end of our $95$~per~cent range. The
results also agree well with \citet{christenson2023} (within the
$95$~per~cent range, and mostly within the $68$~per~cent range), who
employ a narrowband survey to select LAEs around four $z\sim 6$
quasars. The trend of decreasing galaxy density with increasing \lya{}
effective optical depth -- which is expected if the ionising background
is stronger where the galaxy density is larger \citep{davies2018} -- is
very weak; we find that both the most transmissive
($\tauf{} \lesssim 3$) and most opaque ($\tauf{} \gtrsim 5$) sightlines in the
Sherwood-Relics simulation can coincide with galaxy underdensities.

For comparison, the simpler approach based on a halo mass cut of
$M_h > 3\times10^{11}~\msol$ (dotted curves) gives broadly similar results
to the \oiii-emitter modelling for $\tauf < 3$, and is in excellent
agreement with \citetalias{zhu2026}. For $\tauf > 3$, it deviates from
our more detailed \oiii-emitter modelling approach, most notably
around $\tauf \sim 4$. These intermediate opacities are now associated
with galaxy overdensities -- as opposed to underdensities in the
\oiii-emitter model -- and the scatter is significantly
larger. Overall, the simple halo mass cut leads to excellent agreement
with the \citet{christenson2023} data, as well as placing the
\citet{ishimoto2022} points at $\tauf\approx 3$ at the limit of the
$68$~per~cent range (as opposed to the $95$~per~cent range), while
retaining good agreement with the point at $\tauf \approx 4$. This implies
that the haloes hosting the LAEs used in those studies may be more
massive than the haloes hosting \oiii{} emitters
(cf.~\fref{fig:lMh-dist}). However, the large scatter in our mock
survey results implies that -- regardless of the halo mass selection --
the late reionisation model used in Sherwood-Relics, which appears to
be strongly preferred by a wide range of other data \citep[see
e.g.][]{qin2025}, remains consistent with the relation between galaxy
surface density and \lya{} effective optical depth.

In other recent modelling work, \citet{gangolli2025}
found their FlexRT radiative transfer simulations were able to
reproduce the \citet{ishimoto2022} measurements (with the
 $\tauf\approx3$ points falling just outside the $68$~per~cent range), as
well as the \citet{christenson2023} and \citetalias{zhu2026}
measurements for both transmissive and opaque sightlines. However, for
their ``$\dot{N}_{\rm ion}\propto L_{\rm UV}$'' model at the effective
optical depth of the \citetalias{zhu2026} data ($\tauf\approx2.7$), they predict
a larger scatter,
$0.6 \lesssim \Sigma_{\rm LAE}(r_\perp<10~\cmpch)/\langle\Sigma_{\rm LAE}\rangle \lesssim 2.4$, compared to
our
$0.3 \lesssim \Sigma_{\rm \oiii}(r_\perp<10~\cmpch)/\langle\Sigma_{\rm \oiii}\rangle \lesssim 1.6$. Their
median galaxy density marginally favours overdense environments,
whereas our results favour marginally underdense
environments. Using \thesan{}, \citet{garaldi2025} are in good agreement with
 the \citet{ishimoto2022} point at $\tauf \approx 3$, but they are unable
 to reproduce the \citet{ishimoto2022} points at $\tauf \approx 3$. In
addition, they find good agreement between the \citet{christenson2023}
data in the \thesan{} simulation for the transmissive sightlines at $z=5.7$,
although their models appear to undersample at $\tauf>7$, possibly because of the
smaller \thesan{} simulation volume ($L\approx96~{\rm cMpc}$). The preference for slight galaxy
overdensities at $\tauf \approx3$ found by both \citet{gangolli2025} and
\citet{garaldi2025} also suggests they consider more massive haloes in the
calculation of galaxy density, closer to the halo mass cut of
$M_h > 3\times 10^{11}~\msol$ shown in \fref{fig:Sigma-OIII}.

\begin{figure}
  \centering
  \includegraphics[width=\columnwidth]{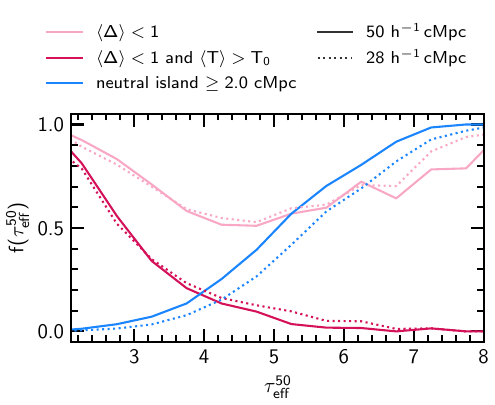}
  \caption{Fraction of segments in each $\tauf$ bin that are, on average, underdense
    ($\langle \Delta \rangle$ < 1, pale pink), and both underdense and hot
    ($\langle T \rangle > T_0$, where $T_0$ is the temperature at $\Delta=1$, pink), where the
    average is calculated over the entire segment. We
    also show the fraction of segments that contain at least one
    neutral island, defined as a contiguous region with
    $x_{\rm \hi}>0.5$ of extent $\geq 2~{\rm cMpc}$ (blue). The
    fraction represents the fraction of segments in each $\tauf$
    bin. We show the fraction for segments of length $50~\cmpch$ (\ie
    the entire segment length used to calculate \tauf, solid) and
    $28~\cmpch$ (\ie the central $28~\cmpch$ corresponding the
    narrowband surveys of LAEs, dotted).}
  \label{fig:f-tauf}
\end{figure}

In \fref{fig:f-tauf} we further explore some of the model behaviour in
\fref{fig:Sigma-OIII}, namely, that both the most transparent ($\tauf
\lesssim 3$) and the most opaque ($\tauf \gtrsim 5$) sightlines exhibit the largest
underdensities of galaxies. We examine this by showing in
\fref{fig:f-tauf} the fraction of segments, $f(\tau_{\rm eff}^{50})$,
that satisfy three sets of constraints:
  \begin{itemize}
  \item where the mean overdensity over the entire segment is less
than $1$ (pale pink curves);
  \item where the mean overdensity over the entire segment is less
than $1$ {\em and} the mean temperature over the entire segment is
greater than the temperature at mean density, $T_0\approx1.2\times10^4~{\rm K}$,
at $z=5.6$ (pink curves)
  \item where the segment contains at least one neutral island,
defined as a contiguous region where $x_{\rm \hi} > 0.5$ for at least
2~cMpc (blue curves).
\end{itemize} We use segments of length either $50~\cmpch$
(corresponding to the length used to calculate \tauf{}, solid curves),
or $28~\cmpch$ (corresponding to the narrowband region to select
galaxies in the centre of the full $50~\cmpch$ sightline, dotted
curves). Both the transparent and opaque sightlines preferentially
correspond to underdense environments, but the transparent sightlines
typically occur in regions that are hotter than the temperature at
mean density, whereas the opaque sightlines correspond to regions that
are colder than the temperature at mean density.  The fraction of
segments containing neutral islands monotonically increases from
$f(\tau_{\rm eff}^{50})\approx0$ at $\tauf \lesssim 3$, to
$f(\tau_{\rm eff}^{50})\approx 1$ for $\tauf \gtrsim 7$. Looking specifically at the
mean values for the 1000 (100) most transparent sightlines, with
$\tauf<2.6~(2.2)$, we find ${\langle T\rangle/T_0}= 1.1~(1.1)$ and the mean
${\langle \Delta\rangle} = 0.79~(0.73)$. Equivalently, for the 1000 (100) most opaque
sightlines with $\tauf> 5.2~(7.9)$, we find
${\langle T\rangle/T_0}= 0.77~(0.60)$ and
${\langle \Delta\rangle} = 0.96~(0.86)$. Taken together, these results suggest that the
most transparent sightlines in our model correspond to underdense
regions that have recently been reionised and have not had time to
cool. By contrast, the most opaque sightlines are colder and have a
much higher likelihood of containing neutral islands, and likely
preferentially reside in regions that are yet to be ionised. This
corresponds to the standard picture of reionisation, where underdense
regions (\ie voids) are typically the last to be reionised.

\section{Conclusions}
\label{sec:conclusions}
We have examined the impact of source and survey modelling on the
connection between high-redshift galaxies and \lya{} forest
transmission. For this purpose we use a model drawn from the
Sherwood-Relics suite of simulations \citep{puchwein2023}, which are
calibrated to \lya{} forest constraints at $z\sim 6$ and are designed to
capture the hydrodynamical effects of inhomogeneous reionisation on
intergalactic gas. We combine this with an empirically calibrated
abundance matching approach that enables us to connect haloes with the
\oiii{}-emitting galaxies observed with {\em JWST}. This model is in
good agreement with recent measurements of the \oiiia{} luminosity
function \citep{matthee2023,sun2023}. Our main findings are as
follows:
\begin{itemize}
\item We find reasonable agreement ($<1$--$2\sigma$) with the clustering of
  \oiii{} emitters recently reported in \citet{huang2026} and
  \citet{eilers2024}, but for a slightly larger minimum luminosity
  $\log_{10}(\loiii/{\rm erg\, s^{-1}})>42.4$ than used by those
  authors. We justify adopting this larger threshold because the
  completeness in \citet{eilers2024} at
  $\log_{10}(\loiii/{\rm erg\, s^{-1}}) = 42.0$ is 40~per~cent, which
  rises to 80~per~cent at
  $\log_{10}(\loiii/{\rm erg\, s^{-1}}) = 42.4$ (see also
  \citealt{pizzati2024}). Any differences between our model and the
  observations at $r_{\perp}\lesssim0.4\rm\,cMpc$ are primarily due to the
  missing one-halo term in our simulation group catalogues.
 
\item We carefully forward model the survey geometry and source
 selection used for recent measurements of the one and two
 dimensional cross-correlation of \oiii{} emitters with the \lya{}
 forest transmission \citep{kakiichi2025,zhu2026}. We find that this
 leads to a slight increase in the average host halo mass
 ($\langle \log_{10} (M_h/\msol)\rangle =10.7$) in our models of the
 galaxy--\lya{} transmission correlation compared the simpler approach
 adopted in our earlier work in \citet{conaboy2025}
 ($\langle \log_{10} (M_h/\msol)\rangle =10.5$). However, this does not change
 the location or magnitude of the observed excess in the one
 dimensional cross-correlation in our models, which peaks at scales
 of $\sim 30\rm\,cMpc$ in the observational data and
 $\sim 20\rm\,cMpc$ in the Sherwood-Relics simulations.
 
\item Our more detailed modelling of the source selection also
  produces a very large scatter in our mock measurements of the
  cross-correlation. Our models are therefore consistent with the
  observational data, within this large scatter, regardless of current
  differences in the predicted and observed shape of the
  cross-correlation. While further exploration of the effect of
  different ionising source models is needed to be certain, the large
  scatter implies that, at present, current galaxy-IGM observations
  may struggle to rule out a broad range of such models.

\item We find good agreement with the latest measurement of the
  relationship between galaxy surface density and \lya{} transmission
  \citep{zhu2026}. Furthermore, in agreement with recent theoretical
  work by \citet{gangolli2025}, we find galaxy underdensities can be
  associated with both the most transparent and most opaque \lya{}
  forest sightlines. This suggests that the relationship between
  galaxies and gas at the tail-end of reionisation is more complex
  than a simple model -- where the ionisation level of the
  intergalactic gas scales proportionally with the galaxy density --
  might imply. We show that in our model the most transparent
  sightlines tend to be underdense and hotter than average, while the
  most opaque sightlines are more likely to correspond to regions that
  have yet to be fully ionised and may contain at least one neutral
  island.
\end{itemize}

We conclude that further theoretical progress will benefit from
simulations performed in larger volumes, and with a wider range of
source models including faint active galactic nuclei
\citep[e.g.][]{asthana2025a}. On the observational front, ongoing
observational programmes such as EIGER \citep{kashino2023} and ASPIRE
\citep{wang2023} that will increase sample sizes will be vital for
further improving our ability to constrain models of reionisation with
the galaxy-IGM connection.

\section*{Acknowledgements}
We thank both anonymous referees, whose constructive comments improved
the quality of this work. LC thanks Koki Kakiichi for sharing the UV
magnitudes and \oiii{} luminosities for the ASPIRE sample, Daichi
Kashino for sharing the latest EIGER measurements of the galaxy--\lya{}
transmission correlation, and Christopher Lovell, Aswin Vijayan, and
Romain Meyer for insightful conversations. The simulations used in
this work were performed using the Joliot Curie supercomputer at the
Tr{\'e}s Grand Centre de Calcul (TGCC) and the Cambridge Service for
Data Driven Discovery (CSD3), part of which is operated by the
University of Cambridge Research Computing on behalf of the STFC DiRAC
HPC Facility (www.dirac.ac.uk). We acknowledge the Partnership for
Advanced Computing in Europe (PRACE) for awarding us time on Joliot
Curie in the 16th call. The DiRAC component of CSD3 was funded by BEIS
capital funding via STFC capital grants ST/P002307/1 and ST/R002452/1
and STFC operations grant ST/R00689X/1. This work also used the
DiRAC@Durham facility managed by the Institute for Computational
Cosmology on behalf of the STFC DiRAC HPC Facility. The equipment was
funded by BEIS capital funding via STFC capital grants ST/P002293/1
and ST/R002371/1, Durham University and STFC operations grant
ST/R000832/1. DiRAC is part of the National e-Infrastructure.  LC and
JSB are supported by STFC consolidated grant ST/X000982/1.  LK
acknowledges the support of a Royal Society University Research
Fellowship (grant number URF$\backslash$R1$\backslash$251793).  Support by ERC Advanced
Grant 320596 `The Emergence of Structure During the Epoch of
Reionization' is gratefully acknowledged. MGH has been supported by
STFC consolidated grants ST/N000927/1 and ST/S000623/1. We thank
Volker Springel for making {\sc p-gadget-3} available. We also thank
Dominique Aubert for sharing the {\sc aton} code, and Philip Parry for
technical support. This work made use of the following open-source
software packages: {\sc cmasher} \citep{vandervelden2020}; {\sc
  matplotlib} \citep{hunter2007}; {\sc numpy} \citep{harris2020}; {\sc
  hmf} \cite{murray2013}; {\sc scipy} \citep{virtanen2020}; and {\sc
  mpmath} \citep{mpmath}.


\section*{Data Availability}
All data and analysis code used in this work are available from the
first author on reasonable request. Further guidance on accessing the
fully publicly available Sherwood-Relics simulation data may also be
found at
\url{https://www.nottingham.ac.uk/astronomy/sherwood-relics/}.

\bibliography{refs}{}

@article{asthana2025a,
  title = {The Impact of Faint {{AGN}} Discovered by {{JWST}} on Reionization},
  author = {Asthana, Shikhar and Haehnelt, Martin G. and Kulkarni, Girish and Bolton, James S. and Gaikwad, Prakash and Keating, Laura C. and Puchwein, Ewald},
  year = 2025,
  month = oct,
  journal = {Monthly Notices of the Royal Astronomical Society},
  volume = {542},
  pages = {2968--2986},
  publisher = {OUP},
  issn = {0035-8711},
  doi = {10.1093/mnras/staf1387},
  urldate = {2026-06-02}
}

@article{aubert2008,
  title = {A Radiative Transfer Scheme for Cosmological Reionization Based on a Local {{Eddington}} Tensor},
  author = {Aubert, Dominique and Teyssier, Romain},
  year = 2008,
  month = jun,
  journal = {Monthly Notices of the Royal Astronomical Society},
  volume = {387},
  pages = {295--307},
  issn = {0035-8711},
  doi = {10.1111/j.1365-2966.2008.13223.x},
  urldate = {2019-06-02}
}

@misc{barlow2003,
  title = {Asymmetric {{Errors}}},
  author = {Barlow, Roger},
  year = 2003,
  month = jan,
  publisher = {arXiv},
  address = {eprint: arXiv:physics/0401042},
  doi = {10.48550/arXiv.physics/0401042},
  urldate = {2025-11-25}
}

@article{basu2026,
  title = {Influence of the Spectral Energy Distribution of Reionization-Era Sources on the {{Lyman-$\alpha$}} Forest},
  author = {Basu, Arghyadeep and Ciardi, Benedetta and Bolton, James S. and Viel, Matteo and Garaldi, Enrico},
  year = 2026,
  month = mar,
  journal = {Monthly Notices of the Royal Astronomical Society},
  volume = {546},
  pages = {stag174},
  publisher = {OUP},
  issn = {0035-8711},
  doi = {10.1093/mnras/stag174},
  urldate = {2026-06-02}
}

@article{becker2018,
  title = {Evidence for {{Large-scale Fluctuations}} in the {{Metagalactic Ionizing Background Near Redshift Six}}},
  author = {Becker, George D. and Davies, Frederick B. and Furlanetto, Steven R. and Malkan, Matthew A. and Boera, Elisa and Douglass, Craig},
  year = 2018,
  month = aug,
  journal = {The Astrophysical Journal},
  volume = {863},
  pages = {92},
  publisher = {IOP},
  issn = {0004-637X},
  doi = {10.3847/1538-4357/aacc73},
  urldate = {2025-12-23}
}

@article{becker2021,
  title = {The Mean Free Path of Ionizing Photons at 5 {$<$} z {$<$} 6: Evidence for Rapid Evolution near Reionization},
  shorttitle = {The Mean Free Path of Ionizing Photons at 5 {$<$} z {$<$} 6},
  author = {Becker, George D. and D'Aloisio, Anson and Christenson, Holly M. and Zhu, Yongda and Worseck, G{\'a}bor and Bolton, James S.},
  year = 2021,
  month = dec,
  journal = {Monthly Notices of the Royal Astronomical Society},
  volume = {508},
  pages = {1853--1869},
  issn = {0035-8711},
  doi = {10.1093/mnras/stab2696},
  urldate = {2022-07-24}
}

@article{becker2024,
  title = {Damping Wing Absorption Associated with a Giant {{Ly}} {$\alpha$} Trough at z {$<$} 6: Direct Evidence for Late-Ending Reionization},
  shorttitle = {Damping Wing Absorption Associated with a Giant {{Ly}} {$\alpha$} Trough at z {$<$} 6},
  author = {Becker, George D. and Bolton, James S. and Zhu, Yongda and Hashemi, Seyedazim},
  year = 2024,
  month = sep,
  journal = {Monthly Notices of the Royal Astronomical Society},
  volume = {533},
  pages = {1525--1540},
  publisher = {OUP},
  issn = {0035-8711},
  doi = {10.1093/mnras/stae1918},
  urldate = {2024-11-06}
}

@article{behroozi2013,
  title = {The {{ROCKSTAR Phase-space Temporal Halo Finder}} and the {{Velocity Offsets}} of {{Cluster Cores}}},
  author = {Behroozi, Peter S. and Wechsler, Risa H. and Wu, Hao-Yi},
  year = 2013,
  month = jan,
  journal = {The Astrophysical Journal},
  volume = {762},
  pages = {109},
  issn = {0004-637X},
  doi = {10.1088/0004-637X/762/2/109},
  urldate = {2019-06-05}
}

@article{bosman2022,
  title = {Hydrogen Reionization Ends by z = 5.3: {{Lyman-$\alpha$}} Optical Depth Measured by the {{XQR-30}} Sample},
  shorttitle = {Hydrogen Reionization Ends by z = 5.3},
  author = {Bosman, Sarah E. I. and Davies, Frederick B. and Becker, George D. and Keating, Laura C. and Davies, Rebecca L. and Zhu, Yongda and Eilers, Anna-Christina and D'Odorico, Valentina and Bian, Fuyan and Bischetti, Manuela and Cristiani, Stefano V. and Fan, Xiaohui and Farina, Emanuele P. and Haehnelt, Martin G. and Hennawi, Joseph F. and Kulkarni, Girish and Mesinger, Andrei and Meyer, Romain A. and Onoue, Masafusa and Pallottini, Andrea and Qin, Yuxiang and {Ryan-Weber}, Emma and Schindler, Jan-Torge and Walter, Fabian and Wang, Feige and Yang, Jinyi},
  year = 2022,
  month = jul,
  journal = {Monthly Notices of the Royal Astronomical Society},
  volume = {514},
  pages = {55--76},
  issn = {0035-8711},
  doi = {10.1093/mnras/stac1046},
  urldate = {2024-04-08}
}

@article{bouwens2008,
  title = {Z \textasciitilde{} 7-10 {{Galaxies}} in the {{HUDF}} and {{GOODS Fields}}: {{UV Luminosity Functions}}},
  shorttitle = {Z \textasciitilde{} 7-10 {{Galaxies}} in the {{HUDF}} and {{GOODS Fields}}},
  author = {Bouwens, Rychard J. and Illingworth, Garth D. and Franx, Marijn and Ford, Holland},
  year = 2008,
  month = oct,
  journal = {The Astrophysical Journal},
  volume = {686},
  pages = {230--250},
  publisher = {IOP},
  issn = {0004-637X},
  doi = {10.1086/590103},
  urldate = {2026-01-08}
}

@article{bouwens2021,
  title = {New {{Determinations}} of the {{UV Luminosity Functions}} from z   9 to 2 {{Show}} a {{Remarkable Consistency}} with {{Halo Growth}} and a {{Constant Star Formation Efficiency}}},
  author = {Bouwens, R. J. and Oesch, P. A. and Stefanon, M. and Illingworth, G. and Labb{\'e}, I. and Reddy, N. and Atek, H. and Montes, M. and Naidu, R. and Nanayakkara, T. and Nelson, E. and Wilkins, S.},
  year = 2021,
  month = aug,
  journal = {The Astronomical Journal},
  volume = {162},
  pages = {47},
  publisher = {IOP},
  issn = {0004-6256},
  doi = {10.3847/1538-3881/abf83e},
  urldate = {2025-08-01}
}

@article{chakraborty2026,
  title = {Probing Reionization-Era Galaxies with {{JWST UV}} Luminosity Functions and Large-Scale Clustering},
  author = {Chakraborty, Anirban and Choudhury, Tirthankar Roy},
  year = 2026,
  month = jan,
  journal = {Journal of Cosmology and Astroparticle Physics},
  volume = {2026},
  pages = {008},
  publisher = {IOP},
  issn = {1475-7516},
  doi = {10.1088/1475-7516/2026/01/008},
  urldate = {2026-06-02}
}

@article{christenson2021,
  title = {Constraints on the {{End}} of {{Reionization}} from the {{Density Fields Surrounding Two Highly Opaque Quasar Sightlines}}},
  author = {Christenson, Holly M. and Becker, George D. and Furlanetto, Steven R. and Davies, Frederick B. and Malkan, Matthew A. and Zhu, Yongda and Boera, Elisa and Trapp, Adam},
  year = 2021,
  month = dec,
  journal = {The Astrophysical Journal},
  volume = {923},
  pages = {87},
  publisher = {IOP},
  issn = {0004-637X},
  doi = {10.3847/1538-4357/ac2a34},
  urldate = {2025-12-23}
}

@article{christenson2023,
  title = {The {{Relationship}} between {{IGM Ly$\alpha$ Opacity}} and {{Galaxy Density}} near the {{End}} of {{Reionization}}},
  author = {Christenson, Holly M. and Becker, George D. and D'Aloisio, Anson and Davies, Frederick B. and Zhu, Yongda and Boera, Elisa and Nasir, Fahad and Furlanetto, Steven R. and Malkan, Matthew A.},
  year = 2023,
  month = oct,
  journal = {The Astrophysical Journal},
  volume = {955},
  pages = {138},
  publisher = {IOP},
  issn = {0004-637X},
  doi = {10.3847/1538-4357/acf450},
  urldate = {2024-04-17}
}

@article{conaboy2025,
  title = {The Connection between High-Redshift Galaxies and {{Lyman}} {$\alpha$} Transmission in the {{Sherwood}}--{{Relics}} Simulations of Patchy Reionization},
  author = {Conaboy, Luke and Bolton, James S. and Keating, Laura C. and Haehnelt, Martin G. and Kulkarni, Girish and Puchwein, Ewald},
  year = 2025,
  month = may,
  journal = {Monthly Notices of the Royal Astronomical Society},
  volume = {539},
  pages = {2790--2805},
  publisher = {OUP},
  issn = {0035-8711},
  doi = {10.1093/mnras/staf648},
  urldate = {2025-05-19}
}

@article{cooray2002,
  title = {Halo Models of Large Scale Structure},
  author = {Cooray, Asantha and Sheth, Ravi},
  year = 2002,
  month = dec,
  journal = {Physics Reports},
  volume = {372},
  pages = {1--129},
  publisher = {Elsevier},
  issn = {0370-1573},
  doi = {10.1016/S0370-1573(02)00276-4},
  urldate = {2026-06-01}
}

@article{cooray2005,
  title = {What Is {{L}}{$\star$}? {{Anatomy}} of the {{Galaxy Luminosity Function}}},
  shorttitle = {What Is {{L}}{$\star$}?},
  author = {Cooray, Asantha and Milosavljevi{\'c}, Milo{\v s}},
  year = 2005,
  month = jun,
  journal = {The Astrophysical Journal},
  volume = {627},
  number = {2},
  pages = {L89},
  publisher = {IOP Publishing},
  issn = {0004-637X},
  doi = {10.1086/432259},
  urldate = {2025-08-11},
  langid = {english}
}

@article{davies2018,
  title = {Quantitative {{Constraints}} on the {{Reionization History}} from the {{IGM Damping Wing Signature}} in {{Two Quasars}} at z {$>$} 7},
  author = {Davies, Frederick B. and Hennawi, Joseph F. and Ba{\~n}ados, Eduardo and Luki{\'c}, Zarija and Decarli, Roberto and Fan, Xiaohui and Farina, Emanuele P. and Mazzucchelli, Chiara and Rix, Hans-Walter and Venemans, Bram P. and Walter, Fabian and Wang, Feige and Yang, Jinyi},
  year = 2018,
  month = sep,
  journal = {The Astrophysical Journal},
  volume = {864},
  pages = {142},
  issn = {0004-637X},
  doi = {10.3847/1538-4357/aad6dc},
  urldate = {2023-02-03}
}

@article{davies2026,
  title = {Updated Dark Pixel Fraction Constraints on Reionization's End from the {{Lyman-series}} Forests of {{XQR}}-30},
  author = {Davies, Frederick B. and Bosman, Sarah E. I. and D'Odorico, Valentina and Campo, Sofia and Mesinger, Andrei and Qin, Yuxiang and Becker, George D. and Ba{\~n}ados, Eduardo and Chen, Huanqing and Cristiani, Stefano and Fan, Xiaohui and Gallerani, Simona and Haehnelt, Martin G. and Keating, Laura C. and Lai, Samuel and {Ryan-Weber}, Emma and Wang, Feige and Yang, Jinyi and Zhu, Yongda},
  year = 2026,
  month = jan,
  journal = {Monthly Notices of the Royal Astronomical Society},
  volume = {545},
  pages = {staf1862},
  publisher = {OUP},
  issn = {0035-8711},
  doi = {10.1093/mnras/staf1862},
  urldate = {2026-02-02}
}

@article{eilers2024,
  title = {{{EIGER}}. {{VI}}. {{The Correlation Function}}, {{Host Halo Mass}}, and {{Duty Cycle}} of {{Luminous Quasars}} at z {$\greaterequivlnt$} 6},
  author = {Eilers, Anna-Christina and Mackenzie, Ruari and Pizzati, Elia and Matthee, Jorryt and Hennawi, Joseph F. and Zhang, Haowen and Bordoloi, Rongmon and Kashino, Daichi and Lilly, Simon J. and Naidu, Rohan P. and Simcoe, Robert A. and Yue, Minghao and Frenk, Carlos S. and Helly, John C. and Schaller, Matthieu and Schaye, Joop},
  year = 2024,
  month = oct,
  journal = {The Astrophysical Journal},
  volume = {974},
  pages = {275},
  publisher = {IOP},
  issn = {0004-637X},
  doi = {10.3847/1538-4357/ad778b},
  urldate = {2025-11-13}
}

@article{gangolli2025,
  title = {On the Correlation between {{Ly$\alpha$}} Forest Opacity and Galaxy Density in Late Reionization Models},
  author = {Gangolli, Nakul and D'Aloisio, Anson and Cain, Christopher and Becker, George D. and Christenson, Holly},
  year = 2025,
  month = mar,
  journal = {Journal of Cosmology and Astroparticle Physics},
  volume = {2025},
  pages = {069},
  publisher = {IOP},
  issn = {1475-7516},
  doi = {10.1088/1475-7516/2025/03/069},
  urldate = {2025-12-23}
}

@article{garaldi2019,
  title = {Constraining the {{Tail End}} of {{Reionization Using Ly$\alpha$ Transmission Spikes}}},
  author = {Garaldi, Enrico and Gnedin, Nickolay Y. and Madau, Piero},
  year = 2019,
  month = may,
  journal = {The Astrophysical Journal},
  volume = {876},
  pages = {31},
  publisher = {IOP},
  issn = {0004-637X},
  doi = {10.3847/1538-4357/ab12dc},
  urldate = {2024-08-29}
}

@article{garaldi2022,
  title = {The {{THESAN}} Project: Properties of the Intergalactic Medium and Its Connection to Reionization-Era Galaxies},
  shorttitle = {The {{THESAN}} Project},
  author = {Garaldi, E. and Kannan, R. and Smith, A. and Springel, V. and Pakmor, R. and Vogelsberger, M. and Hernquist, L.},
  year = 2022,
  month = jun,
  journal = {Monthly Notices of the Royal Astronomical Society},
  volume = {512},
  pages = {4909--4933},
  issn = {0035-8711},
  doi = {10.1093/mnras/stac257},
  urldate = {2024-03-01}
}

@article{garaldi2025,
  title = {The Galaxy-{{IGM}} Connection in {{THESAN}}: The Physics Connecting the {{IGM Lyman-}} {$\alpha$} Opacity and Galaxy Density in the Reionization Epoch},
  shorttitle = {The Galaxy-{{IGM}} Connection in {{THESAN}}},
  author = {Garaldi, Enrico and Bellscheidt, Verena and Smith, Aaron and Kannan, Rahul},
  year = 2025,
  month = aug,
  journal = {The Open Journal of Astrophysics},
  volume = {8},
  pages = {116},
  issn = {2565-6120},
  doi = {10.33232/001c.143245},
  urldate = {2026-02-12}
}

@article{garaldi2025a,
  title = {The Galaxy-{{IGM}} Connection in {{THESAN}}: Observability and Information Content of the Galaxy-{{Lyman-}} {$\alpha$} Cross-Correlation at z {$\geq$} 6},
  shorttitle = {The Galaxy-{{IGM}} Connection in {{THESAN}}},
  author = {Garaldi, Enrico and Bellscheidt, Verena and Smith, Aaron and Kannan, Rahul},
  year = 2025,
  month = dec,
  journal = {The Open Journal of Astrophysics},
  volume = {8},
  pages = {51666},
  issn = {2565-6120},
  doi = {10.33232/001c.151666},
  urldate = {2026-02-12}
}

@article{garcia-vergara2017,
  title = {Strong {{Clustering}} of {{Lyman Break Galaxies}} around {{Luminous Quasars}} at {{Z}} {$\sim$} 4},
  author = {{Garc{\'i}a-Vergara}, Cristina and Hennawi, Joseph F. and Barrientos, L. Felipe and Rix, Hans-Walter},
  year = 2017,
  month = oct,
  journal = {The Astrophysical Journal},
  volume = {848},
  pages = {7},
  publisher = {IOP},
  issn = {0004-637X},
  doi = {10.3847/1538-4357/aa8b69},
  urldate = {2025-11-27}
}

@article{gorce2022,
  title = {Retrieving Cosmological Information from Small-Scale {{CMB}} Foregrounds. {{II}}. {{The}} Kinetic {{Sunyaev Zel}}'dovich Effect},
  author = {Gorce, Ad{\'e}lie and Douspis, Marian and Salvati, Laura},
  year = 2022,
  month = jun,
  journal = {Astronomy and Astrophysics},
  volume = {662},
  pages = {A122},
  publisher = {EDP},
  issn = {0004-6361},
  doi = {10.1051/0004-6361/202243351},
  urldate = {2026-02-02}
}

@inproceedings{greene2016,
  title = {Slitless Spectroscopy with the {{James Webb Space Telescope Near-Infrared Camera}} ({{JWST NIRCam}})},
  booktitle = {Space {{Telescopes}} and {{Instrumentation}} 2016: {{Optical}}, {{Infrared}}, and {{Millimeter Wave}}},
  author = {Greene, Thomas P. and Chu, Laurie and Egami, Eiichi and Hodapp, Klaus W. and Kelly, Douglas M. and Leisenring, Jarron and Rieke, Marcia and Robberto, Massimo and Schlawin, Everett and Stansberry, John},
  year = 2016,
  month = jul,
  volume = {9904},
  pages = {99040E},
  address = {eprint: arXiv:1606.04161},
  doi = {10.1117/12.2231347},
  urldate = {2025-11-12}
}

@article{harris2020,
  title = {Array Programming with {{NumPy}}},
  author = {Harris, Charles R. and Millman, K. Jarrod and {van der Walt}, St{\'e}fan J. and Gommers, Ralf and Virtanen, Pauli and Cournapeau, David and Wieser, Eric and Taylor, Julian and Berg, Sebastian and Smith, Nathaniel J. and Kern, Robert and Picus, Matti and Hoyer, Stephan and {van Kerkwijk}, Marten H. and Brett, Matthew and Haldane, Allan and {del R{\'i}o}, Jaime Fern{\'a}ndez and Wiebe, Mark and Peterson, Pearu and {G{\'e}rard-Marchant}, Pierre and Sheppard, Kevin and Reddy, Tyler and Weckesser, Warren and Abbasi, Hameer and Gohlke, Christoph and Oliphant, Travis E.},
  year = 2020,
  month = sep,
  journal = {Nature},
  volume = {585},
  pages = {357--362},
  issn = {0028-0836},
  doi = {10.1038/s41586-020-2649-2},
  urldate = {2022-06-07}
}

@article{hennawi2006a,
  title = {Binary {{Quasars}} in the {{Sloan Digital Sky Survey}}: {{Evidence}} for {{Excess Clustering}} on {{Small Scales}}},
  shorttitle = {Binary {{Quasars}} in the {{Sloan Digital Sky Survey}}},
  author = {Hennawi, Joseph F. and Strauss, Michael A. and Oguri, Masamune and Inada, Naohisa and Richards, Gordon T. and Pindor, Bartosz and Schneider, Donald P. and Becker, Robert H. and Gregg, Michael D. and Hall, Patrick B. and Johnston, David E. and Fan, Xiaohui and Burles, Scott and Schlegel, David J. and Gunn, James E. and Lupton, Robert H. and Bahcall, Neta A. and Brunner, Robert J. and Brinkmann, Jon},
  year = 2006,
  month = jan,
  journal = {The Astronomical Journal},
  volume = {131},
  pages = {1--23},
  publisher = {IOP},
  issn = {0004-6256},
  doi = {10.1086/498235},
  urldate = {2026-04-20}
}

@misc{huang2026,
  title = {Clustering of Z\textasciitilde 6.6 {{Quasars}} and [{{O III}}] {{Emitters Constrains Host Halo Masses}} and {{Duty Cycles}} in 25 {{ASPIRE Fields}}},
  author = {Huang, Jiamu and Hennawi, Joseph and Pizzati, Elia and Wang, Feige and Yang, Jinyi and Champagne, Jaclyn B. and Fan, Xiaohui and Ba{\~n}ados, Eduardo and Jin, Xiangyu and Kakiichi, Koki and Meyer, Romain A. and Sun, Fengwu and Wu, Yunjing and Zhang, Haowen and Mazzucchelli, Chiara and Eilers, Anna-Christina and Pudoka, Maria and Zhang, Huanian and Schindler, Jan-Torge and Schaller, Matthieu and Schaye, Joop and Snyder, Ben and Kang, Yi and Onorato, Silvia},
  year = 2026,
  month = feb,
  number = {arXiv:2602.04974},
  eprint = {2602.04974},
  primaryclass = {astro-ph},
  publisher = {arXiv},
  doi = {10.48550/arXiv.2602.04974},
  urldate = {2026-02-06},
  archiveprefix = {arXiv}
}

@article{hunter2007,
  title = {Matplotlib: {{A 2D Graphics Environment}}},
  shorttitle = {Matplotlib},
  author = {Hunter, John D.},
  year = 2007,
  month = may,
  journal = {Computing in Science and Engineering},
  volume = {9},
  pages = {90--95},
  doi = {10.1109/MCSE.2007.55},
  urldate = {2022-06-07}
}

@article{ishimoto2022,
  title = {The Physical Origin for Spatially Large Scatter of {{IGM}} Opacity at the End of Reionization: {{The IGM Ly$\alpha$}} Opacity-Galaxy Density Relation},
  shorttitle = {The Physical Origin for Spatially Large Scatter of {{IGM}} Opacity at the End of Reionization},
  author = {Ishimoto, Rikako and Kashikawa, Nobunari and Kashino, Daichi and Ito, Kei and Liang, Yongming and Cai, Zheng and Yoshioka, Takehiro and Okoshi, Katsuya and Misawa, Toru and Onoue, Masafusa and Takeda, Yoshihiro and Uchiyama, Hisakazu},
  year = 2022,
  month = oct,
  journal = {Monthly Notices of the Royal Astronomical Society},
  volume = {515},
  pages = {5914--5926},
  publisher = {OUP},
  issn = {0035-8711},
  doi = {10.1093/mnras/stac1972},
  urldate = {2026-02-02}
}

@article{jin2023,
  title = {({{Nearly}}) {{Model-independent Constraints}} on the {{Neutral Hydrogen Fraction}} in the {{Intergalactic Medium}} at z   5-7 {{Using Dark Pixel Fractions}} in {{Ly$\alpha$}} and {{Ly$\beta$ Forests}}},
  author = {Jin, Xiangyu and Yang, Jinyi and Fan, Xiaohui and Wang, Feige and Ba{\~n}ados, Eduardo and Bian, Fuyan and Davies, Frederick B. and Eilers, Anna-Christina and Farina, Emanuele Paolo and Hennawi, Joseph F. and Pacucci, Fabio and Venemans, Bram and Walter, Fabian},
  year = 2023,
  month = jan,
  journal = {The Astrophysical Journal},
  volume = {942},
  pages = {59},
  issn = {0004-637X},
  doi = {10.3847/1538-4357/aca678},
  urldate = {2023-04-13}
}

@misc{jin2026,
  title = {Probing {{Direct Contributions}} of {{Galaxies}} and {{AGN}} to {{Cosmic Reionization}} in a {{Quasar Field J0226}}+0302 with {{JWST NIRCam}} and {{NIRSpec}}},
  author = {Jin, Xiangyu and Yang, Jinyi and Wang, Feige and Kakiichi, Koki and Fan, Xiaohui and Garaldi, Enrico and Champagne, Jaclyn B. and Becker, George D. and Zhu, Yongda and Wu, Yunjing and Vestergaard, Marianne and Chen, Huanqing and D'Odorico, Valentina and Eilers, Anna-Christina and Huang, Jiamu and Jun, Hyunsung D. and Li, Mingyu and Pudoka, Maria and Tee, Wei Leong and Yue, Minghao and Zhang, Huanian and Zou, Siwei},
  year = 2026,
  month = jun,
  publisher = {arXiv},
  doi = {10.48550/arXiv.2606.14959},
  urldate = {2026-08-06}
}

@article{kakiichi2018,
  title = {The Role of Galaxies and {{AGN}} in Reionizing the {{IGM}} - {{I}}. {{Keck}} Spectroscopy of 5 {$<$} z {$<$} 7 Galaxies in the {{QSO}} Field {{J1148}}+5251},
  author = {Kakiichi, Koki and Ellis, Richard S. and Laporte, Nicolas and Zitrin, Adi and Eilers, Anna-Christina and {Ryan-Weber}, Emma and Meyer, Romain A. and Robertson, Brant and Stark, Daniel P. and Bosman, Sarah E. I.},
  year = 2018,
  month = sep,
  journal = {Monthly Notices of the Royal Astronomical Society},
  volume = {479},
  pages = {43--63},
  issn = {0035-8711},
  doi = {10.1093/mnras/sty1318},
  urldate = {2024-04-04}
}

@article{kakiichi2025,
  title = {{{JWST ASPIRE}}: {{How Did Galaxies Complete Reionization}}? {{Evidence}} for {{Excess IGM Transmission}} around \$\textbraceleft\textbackslash rm [{{O}}\textbackslash,\textbraceleft\textbackslash scriptstyle {{III}}\textbraceright ]\textbraceright\$ {{Emitters}} during {{Reionization}}},
  shorttitle = {{{JWST ASPIRE}}},
  author = {Kakiichi, Koki and Jin, Xiangyu and Wang, Feige and Meyer, Romain A. and Garaldi, Enrico and Bosman, Sarah E. I. and Davies, Frederick B. and Fan, Xiaohui and Trebitsch, Maxime and Yang, Jinyi and Ba{\~n}ados, Eduardo and Champagne, Jaclyn B. and Eilers, Anna-Christina and Hennawi, Joseph F. and Sun, Fengwu and Wu, Yunjing and Zou, Siwei and Kannan, Rahul and Smith, Aaron and Becker, George D. and D'Odorico, Valentina and Connor, Thomas and Liu, Weizhe and Protu{\v s}ov{\'a}, Klaudia and Walter, Fabian and Zhang, Huanian},
  year = 2025,
  month = mar,
  journal = {arXiv e-prints},
  eprint = {2503.07074},
  primaryclass = {astro-ph},
  pages = {arXiv:2503.07074},
  doi = {10.48550/arXiv.2503.07074},
  urldate = {2025-03-11},
  archiveprefix = {arXiv}
}

@article{kashino2020,
  title = {Evidence for a {{Highly Opaque Large-scale Galaxy Void}} at the {{End}} of {{Reionization}}},
  author = {Kashino, Daichi and Lilly, Simon J. and Shibuya, Takatoshi and Ouchi, Masami and Kashikawa, Nobunari},
  year = 2020,
  month = jan,
  journal = {The Astrophysical Journal},
  volume = {888},
  pages = {6},
  publisher = {IOP},
  issn = {0004-637X},
  doi = {10.3847/1538-4357/ab5a7d},
  urldate = {2025-12-23}
}

@article{kashino2023,
  title = {{{EIGER}}. {{I}}. {{A Large Sample}} of [{{O III}}]-Emitting {{Galaxies}} at 5.3 {$<$} z {$<$} 6.9 and {{Direct Evidence}} for {{Local Reionization}} by {{Galaxies}}},
  author = {Kashino, Daichi and Lilly, Simon J. and Matthee, Jorryt and Eilers, Anna-Christina and Mackenzie, Ruari and Bordoloi, Rongmon and Simcoe, Robert A.},
  year = 2023,
  month = jun,
  journal = {The Astrophysical Journal},
  volume = {950},
  pages = {66},
  issn = {0004-637X},
  doi = {10.3847/1538-4357/acc588},
  urldate = {2024-04-04}
}

@article{kashino2026,
  title = {{{EIGER}}. {{VII}}. {{The Evolving Relationship}} between {{Galaxies}} and the {{Intergalactic Medium}} in the {{Final Stages}} of {{Reionization}}},
  author = {Kashino, Daichi and Lilly, Simon J. and Matthee, Jorryt and Mackenzie, Ruari and Eilers, Anna-Christina and Bordoloi, Rongmon and Simcoe, Robert A. and Naidu, Rohan P. and Yue, Minghao and Liu, Bin},
  year = 2026,
  month = feb,
  journal = {The Astrophysical Journal},
  volume = {997},
  pages = {280},
  publisher = {IOP},
  issn = {0004-637X},
  doi = {10.3847/1538-4357/ae2799},
  urldate = {2026-02-13}
}

@article{keating2020,
  title = {Constraining the Second Half of Reionization with the {{Ly}} {$\beta$} Forest},
  author = {Keating, Laura C. and Kulkarni, Girish and Haehnelt, Martin G. and Chardin, Jonathan and Aubert, Dominique},
  year = 2020,
  month = sep,
  journal = {Monthly Notices of the Royal Astronomical Society},
  volume = {497},
  pages = {906--915},
  publisher = {OUP},
  issn = {0035-8711},
  doi = {10.1093/mnras/staa1909},
  urldate = {2024-11-06}
}

@article{keating2020a,
  title = {Long Troughs in the {{Lyman-$\alpha$}} Forest below Redshift 6 Due to Islands of Neutral Hydrogen},
  author = {Keating, Laura C. and Weinberger, Lewis H. and Kulkarni, Girish and Haehnelt, Martin G. and Chardin, Jonathan and Aubert, Dominique},
  year = 2020,
  month = jan,
  journal = {Monthly Notices of the Royal Astronomical Society},
  volume = {491},
  pages = {1736--1745},
  publisher = {OUP},
  issn = {0035-8711},
  doi = {10.1093/mnras/stz3083},
  urldate = {2024-04-17}
}

@article{kulkarni2019,
  title = {Large {{Ly}} {$\alpha$} Opacity Fluctuations and Low {{CMB}} {$\tau$} in Models of Late Reionization with Large Islands of Neutral Hydrogen Extending to z {$<$} 5.5},
  author = {Kulkarni, Girish and Keating, Laura C. and Haehnelt, Martin G. and Bosman, Sarah E. I. and Puchwein, Ewald and Chardin, Jonathan and Aubert, Dominique},
  year = 2019,
  month = may,
  journal = {Monthly Notices of the Royal Astronomical Society},
  volume = {485},
  pages = {L24-L28},
  publisher = {OUP},
  issn = {0035-8711},
  doi = {10.1093/mnrasl/slz025},
  urldate = {2024-08-01}
}

@article{landy1993,
  title = {Bias and {{Variance}} of {{Angular Correlation Functions}}},
  author = {Landy, Stephen D. and Szalay, Alexander S.},
  year = 1993,
  month = jul,
  journal = {The Astrophysical Journal},
  volume = {412},
  pages = {64},
  publisher = {IOP},
  issn = {0004-637X},
  doi = {10.1086/172900},
  urldate = {2025-11-27}
}

@article{lee2009,
  title = {Mapping the {{Dark Matter}} from {{UV Light}} at {{High Redshift}}: {{An Empirical Approach}} to {{Understand Galaxy Statistics}}},
  shorttitle = {Mapping the {{Dark Matter}} from {{UV Light}} at {{High Redshift}}},
  author = {Lee, Kyoung-Soo and Giavalisco, Mauro and Conroy, Charlie and Wechsler, Risa H. and Ferguson, Henry C. and Somerville, Rachel S. and Dickinson, Mark E. and Urry, Claudia M.},
  year = 2009,
  month = apr,
  journal = {The Astrophysical Journal},
  volume = {695},
  pages = {368--390},
  issn = {0004-637X},
  doi = {10.1088/0004-637X/695/1/368},
  urldate = {2025-09-03}
}

@article{maitra2025,
  title = {The {{Lyman}} {$\alpha$} Emitter Bispectrum as a Probe of Reionization Morphology},
  author = {Maitra, Soumak and Kulkarni, Girish and Asthana, Shikhar and Bolton, James S. and Haehnelt, Martin G. and Keating, Laura},
  year = 2025,
  month = sep,
  journal = {Monthly Notices of the Royal Astronomical Society},
  volume = {542},
  pages = {486--507},
  issn = {0035-8711},
  doi = {10.1093/mnras/staf1262},
  urldate = {2025-09-03}
}

@article{matthee2023,
  title = {{{EIGER}}. {{II}}. {{First Spectroscopic Characterization}} of the {{Young Stars}} and {{Ionized Gas Associated}} with {{Strong H$\beta$}} and [{{O III}}] {{Line Emission}} in {{Galaxies}} at z = 5-7 with {{JWST}}},
  author = {Matthee, Jorryt and Mackenzie, Ruari and Simcoe, Robert A. and Kashino, Daichi and Lilly, Simon J. and Bordoloi, Rongmon and Eilers, Anna-Christina},
  year = 2023,
  month = jun,
  journal = {The Astrophysical Journal},
  volume = {950},
  pages = {67},
  publisher = {IOP},
  issn = {0004-637X},
  doi = {10.3847/1538-4357/acc846},
  urldate = {2025-07-21}
}

@article{mcgreer2015,
  title = {Model-Independent Evidence in Favour of an End to Reionization by z {$\approx$} 6},
  author = {McGreer, Ian D. and Mesinger, Andrei and D'Odorico, Valentina},
  year = 2015,
  month = feb,
  journal = {Monthly Notices of the Royal Astronomical Society},
  volume = {447},
  pages = {499--505},
  issn = {0035-8711},
  doi = {10.1093/mnras/stu2449},
  urldate = {2023-04-13}
}

@article{meyer2019,
  title = {The Role of Galaxies and {{AGNs}} in Reionizing the {{IGM}} - {{II}}. {{Metal-tracing}} the Faint Sources of Reionization at 5 {$\lessequivlnt$} z {$\lessequivlnt$} 6},
  author = {Meyer, Romain A. and Bosman, Sarah E. I. and Kakiichi, Koki and Ellis, Richard S.},
  year = 2019,
  month = feb,
  journal = {Monthly Notices of the Royal Astronomical Society},
  volume = {483},
  pages = {19--37},
  issn = {0035-8711},
  doi = {10.1093/mnras/sty2954},
  urldate = {2024-04-04}
}

@article{meyer2020,
  title = {The Role of Galaxies and {{AGN}} in Reionizing the {{IGM}} - {{III}}. {{IGM-galaxy}} Cross-Correlations at z {$\sim$} 6 from Eight Quasar Fields with {{DEIMOS}} and {{MUSE}}},
  author = {Meyer, Romain A. and Kakiichi, Koki and Bosman, Sarah E. I. and Ellis, Richard S. and Laporte, Nicolas and Robertson, Brant E. and {Ryan-Weber}, Emma V. and Mawatari, Ken and Zitrin, Adi},
  year = 2020,
  month = may,
  journal = {Monthly Notices of the Royal Astronomical Society},
  volume = {494},
  pages = {1560--1578},
  issn = {0035-8711},
  doi = {10.1093/mnras/staa746},
  urldate = {2024-04-04}
}

@article{meyer2024,
  title = {{{JWST FRESCO}}: A Comprehensive Census of {{H}}\,{$\beta~$}+~[{{O}} Iii] Emitters at 6.8 {$<$} z {$<$} 9.0 in the {{GOODS}} Fields},
  shorttitle = {{{JWST FRESCO}}},
  author = {Meyer, R A and Oesch, P A and Giovinazzo, E and Weibel, A and Brammer, G and Matthee, J and Naidu, R P and Bouwens, R J and Chisholm, J and {Covelo-Paz}, A and Fudamoto, Y and Maseda, M and Nelson, E and Shivaei, I and Xiao, M and {Herard-Demanche}, T and Illingworth, G D and Kerutt, J and Kramarenko, I and Labbe, I and Leonova, E and Magee, D and Matharu, J and Prieto~Lyon, G and Reddy, N and Schaerer, D and Shapley, A and Stefanon, M and Wozniak, M A and Wuyts, S},
  year = 2024,
  month = nov,
  journal = {Monthly Notices of the Royal Astronomical Society},
  volume = {535},
  number = {1},
  pages = {1067--1094},
  issn = {0035-8711},
  doi = {10.1093/mnras/stae2353},
  urldate = {2025-12-23}
}

@misc{meyer2025,
  title = {{{JWST COSMOS-3D}}: {{Spectroscopic Census}} and {{Luminosity Function}} of [{{O III}}] {{Emitters}} at 6.75},
  shorttitle = {{{JWST COSMOS-3D}}},
  author = {Meyer, Romain A. and Wang, Feige and Kakiichi, Koki and Brammer, Gabe and Champagne, Jackie and Jurk, Katharina and Li, Zihao and Li, Zijian and Musin, Marat and Satyavolu, Sindhu and Schindler, Jan-Torge and Shuntov, Marko and Xu, Yi and Zou, Siwei and Bian, Fuyan and Casey, Caitlin and Egami, Eiichi and Fan, Xiaohui and Jiang, Danyang and Laporte, Nicolas and Liu, Weizhe and Oesch, Pascal and Tasca, Lidia and Yang, Jinyi and Zhang, Zijian and Akins, Hollis and Cai, Zheng and Coulter, Dave A. and Huang, Jiamu and Li, Mingyu and Liu, Weizhe and Liang, Yongming and Jin, Xiangyu and Kartaltepe, Jeyhan and Matharu, Jasleen and Pudoka, Maria and Tee, Wei-Leong and Witten, Callum and Zhang, Haowen and Zhu, Yongda},
  year = 2025,
  month = oct,
  publisher = {arXiv},
  doi = {10.48550/arXiv.2510.11373},
  urldate = {2026-01-14}
}

@misc{mpmath,
  title = {Mpmath: A {{Python}} Library for Arbitrary-Precision Floating-Point Arithmetic (Version 1.3.0)},
  author = {{The mpmath development team}},
  year = 2023
}

@article{murray2013,
  title = {{{HMFcalc}}: {{An}} Online Tool for Calculating Dark Matter Halo Mass Functions},
  shorttitle = {{{HMFcalc}}},
  author = {Murray, S. G. and Power, C. and Robotham, A. S. G.},
  year = 2013,
  month = nov,
  journal = {Astronomy and Computing},
  volume = {3},
  pages = {23},
  issn = {2213-1337},
  doi = {10.1016/j.ascom.2013.11.001},
  urldate = {2023-03-13},
  langid = {english}
}

@article{oke1983,
  title = {Secondary Standard Stars for Absolute Spectrophotometry.},
  author = {Oke, J. B. and Gunn, J. E.},
  year = 1983,
  month = mar,
  journal = {The Astrophysical Journal},
  volume = {266},
  pages = {713--717},
  publisher = {IOP},
  issn = {0004-637X},
  doi = {10.1086/160817},
  urldate = {2025-11-18}
}

@article{pizzati2024,
  title = {A Unified Model for the Clustering of Quasars and Galaxies at z {$\approx$} 6},
  author = {Pizzati, Elia and Hennawi, Joseph F. and Schaye, Joop and Schaller, Matthieu and Eilers, Anna-Christina and Wang, Feige and Frenk, Carlos S. and Elbers, Willem and Helly, John C. and Mackenzie, Ruari and Matthee, Jorryt and Bordoloi, Rongmon and Kashino, Daichi and Naidu, Rohan P. and Yue, Minghao},
  year = 2024,
  month = nov,
  journal = {Monthly Notices of the Royal Astronomical Society},
  volume = {534},
  pages = {3155--3175},
  publisher = {OUP},
  issn = {0035-8711},
  doi = {10.1093/mnras/stae2307},
  urldate = {2024-11-25}
}

@article{planckcollaboration2014,
  title = {{\emph{Planck}} 2013 Results. {{XVI}}. {{Cosmological}} Parameters},
  author = {{Planck Collaboration} and Ade, P. A. R. and Aghanim, N. and {Armitage-Caplan}, C. and Arnaud, M. and Ashdown, M. and {Atrio-Barandela}, F. and Aumont, J. and Baccigalupi, C. and Banday, A. J. and Barreiro, R. B. and Bartlett, J. G. and Battaner, E. and Benabed, K. and Beno{\^i}t, A. and {Benoit-L{\'e}vy}, A. and Bernard, J.-P. and Bersanelli, M. and Bielewicz, P. and Bobin, J. and Bock, J. J. and Bonaldi, A. and Bond, J. R. and Borrill, J. and Bouchet, F. R. and Bridges, M. and Bucher, M. and Burigana, C. and Butler, R. C. and Calabrese, E. and Cappellini, B. and Cardoso, J.-F. and Catalano, A. and Challinor, A. and Chamballu, A. and Chary, R.-R. and Chen, X. and Chiang, H. C. and Chiang, L.-Y and Christensen, P. R. and Church, S. and Clements, D. L. and Colombi, S. and Colombo, L. P. L. and Couchot, F. and Coulais, A. and Crill, B. P. and Curto, A. and Cuttaia, F. and Danese, L. and Davies, R. D. and Davis, R. J. and {de Bernardis}, P. and {de Rosa}, A. and {de Zotti}, G. and Delabrouille, J. and Delouis, J.-M. and D{\'e}sert, F.-X. and Dickinson, C. and Diego, J. M. and Dolag, K. and Dole, H. and Donzelli, S. and Dor{\'e}, O. and Douspis, M. and Dunkley, J. and Dupac, X. and Efstathiou, G. and Elsner, F. and En{\ss}lin, T. A. and Eriksen, H. K. and Finelli, F. and Forni, O. and Frailis, M. and Fraisse, A. A. and Franceschi, E. and Gaier, T. C. and Galeotta, S. and Galli, S. and Ganga, K. and Giard, M. and Giardino, G. and {Giraud-H{\'e}raud}, Y. and Gjerl{\o}w, E. and {Gonz{\'a}lez-Nuevo}, J. and G{\'o}rski, K. M. and Gratton, S. and Gregorio, A. and Gruppuso, A. and Gudmundsson, J. E. and Haissinski, J. and Hamann, J. and Hansen, F. K. and Hanson, D. and Harrison, D. and {Henrot-Versill{\'e}}, S. and {Hern{\'a}ndez-Monteagudo}, C. and Herranz, D. and Hildebrandt, S. R. and Hivon, E. and Hobson, M. and Holmes, W. A. and Hornstrup, A. and Hou, Z. and Hovest, W. and Huffenberger, K. M. and Jaffe, A. H. and Jaffe, T. R. and Jewell, J. and Jones, W. C. and Juvela, M. and Keih{\"a}nen, E. and Keskitalo, R. and Kisner, T. S. and Kneissl, R. and Knoche, J. and Knox, L. and Kunz, M. and {Kurki-Suonio}, H. and Lagache, G. and L{\"a}hteenm{\"a}ki, A. and Lamarre, J.-M. and Lasenby, A. and Lattanzi, M. and Laureijs, R. J. and Lawrence, C. R. and Leach, S. and Leahy, J. P. and Leonardi, R. and {Le{\'o}n-Tavares}, J. and Lesgourgues, J. and Lewis, A. and Liguori, M. and Lilje, P. B. and {Linden-V{\o}rnle}, M. and {L{\'o}pez-Caniego}, M. and Lubin, P. M. and {Mac{\'i}as-P{\'e}rez}, J. F. and Maffei, B. and Maino, D. and Mandolesi, N. and Maris, M. and Marshall, D. J. and Martin, P. G. and {Mart{\'i}nez-Gonz{\'a}lez}, E. and Masi, S. and Massardi, M. and Matarrese, S. and Matthai, F. and Mazzotta, P. and Meinhold, P. R. and Melchiorri, A. and Melin, J.-B. and Mendes, L. and Menegoni, E. and Mennella, A. and Migliaccio, M. and Millea, M. and Mitra, S. and {Miville-Desch{\^e}nes}, M.-A. and Moneti, A. and Montier, L. and Morgante, G. and Mortlock, D. and Moss, A. and Munshi, D. and Murphy, J. A. and Naselsky, P. and Nati, F. and Natoli, P. and Netterfield, C. B. and {N{\o}rgaard-Nielsen}, H. U. and Noviello, F. and Novikov, D. and Novikov, I. and O'Dwyer, I. J. and Osborne, S. and Oxborrow, C. A. and Paci, F. and Pagano, L. and Pajot, F. and Paladini, R. and Paoletti, D. and Partridge, B. and Pasian, F. and Patanchon, G. and Pearson, D. and Pearson, T. J. and Peiris, H. V. and Perdereau, O. and Perotto, L. and Perrotta, F. and Pettorino, V. and Piacentini, F. and Piat, M. and Pierpaoli, E. and Pietrobon, D. and Plaszczynski, S. and Platania, P. and Pointecouteau, E. and Polenta, G. and Ponthieu, N. and Popa, L. and Poutanen, T. and Pratt, G. W. and Pr{\'e}zeau, G. and Prunet, S. and Puget, J.-L. and Rachen, J. P. and Reach, W. T. and Rebolo, R. and Reinecke, M. and Remazeilles, M. and Renault, C. and Ricciardi, S. and Riller, T. and Ristorcelli, I. and Rocha, G. and Rosset, C. and Roudier, G. and {Rowan-Robinson}, M. and {Rubi{\~n}o-Mart{\'i}n}, J. A. and Rusholme, B. and Sandri, M. and Santos, D. and Savelainen, M. and Savini, G. and Scott, D. and Seiffert, M. D. and Shellard, E. P. S. and Spencer, L. D. and Starck, J.-L. and Stolyarov, V. and Stompor, R. and Sudiwala, R. and Sunyaev, R. and Sureau, F. and Sutton, D. and {Suur-Uski}, A.-S. and Sygnet, J.-F. and Tauber, J. A. and Tavagnacco, D. and Terenzi, L. and Toffolatti, L. and Tomasi, M. and Tristram, M. and Tucci, M. and Tuovinen, J. and T{\"u}rler, M. and Umana, G. and Valenziano, L. and Valiviita, J. and Van Tent, B. and Vielva, P. and Villa, F. and Vittorio, N. and Wade, L. A. and Wandelt, B. D. and Wehus, I. K. and White, M. and White, S. D. M. and Wilkinson, A. and Yvon, D. and Zacchei, A. and Zonca, A.},
  year = 2014,
  month = nov,
  journal = {Astronomy \& Astrophysics},
  volume = {571},
  pages = {A16},
  issn = {0004-6361, 1432-0746},
  doi = {10.1051/0004-6361/201321591},
  urldate = {2022-07-20},
  langid = {english}
}

@article{planckcollaboration2020,
  title = {{\emph{Planck}} 2018 Results: {{VI}}. {{Cosmological}} Parameters},
  shorttitle = {{\emph{Planck}} 2018 Results},
  author = {{Planck Collaboration} and Aghanim, N. and Akrami, Y. and Ashdown, M. and Aumont, J. and Baccigalupi, C. and Ballardini, M. and Banday, A. J. and Barreiro, R. B. and Bartolo, N. and Basak, S. and Battye, R. and Benabed, K. and Bernard, J.-P. and Bersanelli, M. and Bielewicz, P. and Bock, J. J. and Bond, J. R. and Borrill, J. and Bouchet, F. R. and Boulanger, F. and Bucher, M. and Burigana, C. and Butler, R. C. and Calabrese, E. and Cardoso, J.-F. and Carron, J. and Challinor, A. and Chiang, H. C. and Chluba, J. and Colombo, L. P. L. and Combet, C. and Contreras, D. and Crill, B. P. and Cuttaia, F. and {de Bernardis}, P. and {de Zotti}, G. and Delabrouille, J. and Delouis, J.-M. and Di Valentino, E. and Diego, J. M. and Dor{\'e}, O. and Douspis, M. and Ducout, A. and Dupac, X. and Dusini, S. and Efstathiou, G. and Elsner, F. and En{\ss}lin, T. A. and Eriksen, H. K. and Fantaye, Y. and Farhang, M. and Fergusson, J. and {Fernandez-Cobos}, R. and Finelli, F. and Forastieri, F. and Frailis, M. and Fraisse, A. A. and Franceschi, E. and Frolov, A. and Galeotta, S. and Galli, S. and Ganga, K. and {G{\'e}nova-Santos}, R. T. and Gerbino, M. and Ghosh, T. and {Gonz{\'a}lez-Nuevo}, J. and G{\'o}rski, K. M. and Gratton, S. and Gruppuso, A. and Gudmundsson, J. E. and Hamann, J. and Handley, W. and Hansen, F. K. and Herranz, D. and Hildebrandt, S. R. and Hivon, E. and Huang, Z. and Jaffe, A. H. and Jones, W. C. and Karakci, A. and Keih{\"a}nen, E. and Keskitalo, R. and Kiiveri, K. and Kim, J. and Kisner, T. S. and Knox, L. and Krachmalnicoff, N. and Kunz, M. and {Kurki-Suonio}, H. and Lagache, G. and Lamarre, J.-M. and Lasenby, A. and Lattanzi, M. and Lawrence, C. R. and Le Jeune, M. and Lemos, P. and Lesgourgues, J. and Levrier, F. and Lewis, A. and Liguori, M. and Lilje, P. B. and Lilley, M. and Lindholm, V. and {L{\'o}pez-Caniego}, M. and Lubin, P. M. and Ma, Y.-Z. and {Mac{\'i}as-P{\'e}rez}, J. F. and Maggio, G. and Maino, D. and Mandolesi, N. and Mangilli, A. and {Marcos-Caballero}, A. and Maris, M. and Martin, P. G. and Martinelli, M. and {Mart{\'i}nez-Gonz{\'a}lez}, E. and Matarrese, S. and Mauri, N. and McEwen, J. D. and Meinhold, P. R. and Melchiorri, A. and Mennella, A. and Migliaccio, M. and Millea, M. and Mitra, S. and {Miville-Desch{\^e}nes}, M.-A. and Molinari, D. and Montier, L. and Morgante, G. and Moss, A. and Natoli, P. and {N{\o}rgaard-Nielsen}, H. U. and Pagano, L. and Paoletti, D. and Partridge, B. and Patanchon, G. and Peiris, H. V. and Perrotta, F. and Pettorino, V. and Piacentini, F. and Polastri, L. and Polenta, G. and Puget, J.-L. and Rachen, J. P. and Reinecke, M. and Remazeilles, M. and Renzi, A. and Rocha, G. and Rosset, C. and Roudier, G. and {Rubi{\~n}o-Mart{\'i}n}, J. A. and {Ruiz-Granados}, B. and Salvati, L. and Sandri, M. and Savelainen, M. and Scott, D. and Shellard, E. P. S. and Sirignano, C. and Sirri, G. and Spencer, L. D. and Sunyaev, R. and {Suur-Uski}, A.-S. and Tauber, J. A. and Tavagnacco, D. and Tenti, M. and Toffolatti, L. and Tomasi, M. and Trombetti, T. and Valenziano, L. and Valiviita, J. and Van Tent, B. and Vibert, L. and Vielva, P. and Villa, F. and Vittorio, N. and Wandelt, B. D. and Wehus, I. K. and White, M. and White, S. D. M. and Zacchei, A. and Zonca, A.},
  year = 2020,
  month = sep,
  journal = {Astronomy \& Astrophysics},
  volume = {641},
  pages = {A6},
  issn = {0004-6361, 1432-0746},
  doi = {10.1051/0004-6361/201833910},
  urldate = {2022-07-20},
  langid = {english}
}

@article{puchwein2023,
  title = {The {{Sherwood-Relics}} Simulations: Overview and Impact of Patchy Reionization and Pressure Smoothing on the Intergalactic Medium},
  shorttitle = {The {{Sherwood-Relics}} Simulations},
  author = {Puchwein, Ewald and Bolton, James S. and Keating, Laura C. and Molaro, Margherita and Gaikwad, Prakash and Kulkarni, Girish and Haehnelt, Martin G. and Ir{\v s}i{\v c}, Vid and {\v S}oltinsk{\'y}, Tom{\'a}{\v s} and Viel, Matteo and Aubert, Dominique and Becker, George D. and Meiksin, Avery},
  year = 2023,
  month = mar,
  journal = {Monthly Notices of the Royal Astronomical Society},
  volume = {519},
  pages = {6162--6183},
  issn = {0035-8711},
  doi = {10.1093/mnras/stac3761},
  urldate = {2024-04-04}
}

@article{qin2025,
  title = {Percent-Level Timing of Reionisation: {{Self-consistent}}, Implicit-Likelihood Inference from {{XQR-30}}+ {{Ly$\alpha$}} Forest Data},
  shorttitle = {Percent-Level Timing of Reionisation},
  author = {Qin, Yuxiang and Mesinger, Andrei and Prelogovi{\'c}, David and Becker, George and Bischetti, Manuela and Bosman, Sarah and Davies, Frederick and D'Odorico, Valentina and Gaikwad, Prakash and Haehnelt, Martin and Keating, Laura and Lai, Samuel and {Ryan-Weber}, Emma and Satyavolu, Sindhu and Walter, Fabian and Zhu, Yongda},
  year = 2025,
  month = apr,
  journal = {Publications of the Astronomical Society of Australia},
  volume = {42},
  pages = {e049},
  issn = {1323-3580},
  doi = {10.1017/pasa.2025.35},
  urldate = {2026-05-22}
}

@article{rosdahl2022,
  title = {{{LyC}} Escape from {{SPHINX}} Galaxies in the {{Epoch}} of {{Reionization}}},
  author = {Rosdahl, Joakim and Blaizot, J{\'e}r{\'e}my and Katz, Harley and Kimm, Taysun and Garel, Thibault and Haehnelt, Martin and Keating, Laura C. and {Martin-Alvarez}, Sergio and {Michel-Dansac}, L{\'e}o and Ocvirk, Pierre},
  year = 2022,
  month = sep,
  journal = {Monthly Notices of the Royal Astronomical Society},
  volume = {515},
  pages = {2386--2414},
  issn = {0035-8711},
  doi = {10.1093/mnras/stac1942},
  urldate = {2025-09-03}
}

@misc{sarkar2026,
  title = {Reconciling Large-Scale {{Lyman-}}\${$\alpha\$$} Correlations with the {{SCRIPT Semi-numerical Model}}},
  author = {Sarkar, Saptarshi and Choudhury, Tirthankar Roy},
  year = 2026,
  month = jun,
  publisher = {arXiv},
  doi = {10.48550/arXiv.2606.08368},
  urldate = {2026-08-11}
}

@article{satyavolu2024,
  title = {Robustness of Direct Measurements of the Mean Free Path of Ionizing Photons in the Epoch of Reionization},
  author = {Satyavolu, Sindhu and Kulkarni, Girish and Keating, Laura C. and Haehnelt, Martin G.},
  year = 2024,
  month = sep,
  journal = {Monthly Notices of the Royal Astronomical Society},
  volume = {533},
  pages = {676--686},
  publisher = {OUP},
  issn = {0035-8711},
  doi = {10.1093/mnras/stae1717},
  urldate = {2024-11-25}
}

@article{sheth2001,
  title = {Ellipsoidal Collapse and an Improved Model for the Number and Spatial Distribution of Dark Matter Haloes},
  author = {Sheth, Ravi K. and Mo, H. J. and Tormen, Giuseppe},
  year = 2001,
  month = may,
  journal = {Monthly Notices of the Royal Astronomical Society},
  volume = {323},
  pages = {1--12},
  issn = {0035-8711},
  doi = {10.1046/j.1365-8711.2001.04006.x},
  urldate = {2022-07-27}
}

@article{sinha2020,
  title = {{{CORRFUNC}} - a Suite of Blazing Fast Correlation Functions on the {{CPU}}},
  author = {Sinha, Manodeep and Garrison, Lehman H.},
  year = 2020,
  month = jan,
  journal = {Monthly Notices of the Royal Astronomical Society},
  volume = {491},
  pages = {3022--3041},
  publisher = {OUP},
  issn = {0035-8711},
  doi = {10.1093/mnras/stz3157},
  urldate = {2025-11-27}
}

@article{spina2024,
  title = {Damping Wings in the {{Lyman-$\alpha$}} Forest: {{A}} Model-Independent Measurement of the Neutral Fraction at 5.4 {$<$} z {$<$} 6.1},
  shorttitle = {Damping Wings in the {{Lyman-$\alpha$}} Forest},
  author = {Spina, Benedetta and Bosman, Sarah E. I. and Davies, Frederick B. and Gaikwad, Prakash and Zhu, Yongda},
  year = 2024,
  month = aug,
  journal = {Astronomy and Astrophysics},
  volume = {688},
  pages = {L26},
  issn = {0004-6361},
  doi = {10.1051/0004-6361/202450798},
  urldate = {2024-11-19}
}

@article{spina2026,
  title = {Measuring the Intergalactic Medium Correlation Length at 5 {$<$} z {$<$} 6.1: {{A}} Fast Change at the End of Reionization},
  shorttitle = {Measuring the Intergalactic Medium Correlation Length at 5 {$<$} z {$<$} 6.1},
  author = {Spina, Benedetta and Bosman, Sarah E. I. and Maity, Barun and Davies, Frederick B.},
  year = 2026,
  month = feb,
  journal = {Astronomy and Astrophysics},
  volume = {706},
  pages = {A273},
  publisher = {EDP},
  issn = {0004-6361},
  doi = {10.1051/0004-6361/202556442},
  urldate = {2026-08-11}
}

@article{springel2005,
  title = {The Cosmological Simulation Code {{GADGET-2}}},
  author = {Springel, Volker},
  year = 2005,
  month = dec,
  journal = {Monthly Notices of the Royal Astronomical Society},
  volume = {364},
  pages = {1105--1134},
  issn = {0035-8711},
  doi = {10.1111/j.1365-2966.2005.09655.x},
  urldate = {2019-06-04}
}

@article{storey2000,
  title = {Theoretical Values for the [{{OIII}}] 5007/4959 Line-Intensity Ratio and Homologous Cases},
  author = {Storey, P. J. and Zeippen, C. J.},
  year = 2000,
  month = mar,
  journal = {Monthly Notices of the Royal Astronomical Society},
  volume = {312},
  pages = {813--816},
  publisher = {OUP},
  issn = {0035-8711},
  doi = {10.1046/j.1365-8711.2000.03184.x},
  urldate = {2025-11-10}
}

@article{sun2023,
  title = {First {{Sample}} of {{H$\alpha$}}+[{{O III}}]{$\lambda$}5007 {{Line Emitters}} at z {$>$} 6 {{Through JWST}}/{{NIRCam Slitless Spectroscopy}}: {{Physical Properties}} and {{Line-luminosity Functions}}},
  shorttitle = {First {{Sample}} of {{H$\alpha$}}+[{{O III}}]{$\lambda$}5007 {{Line Emitters}} at z {$>$} 6 {{Through JWST}}/{{NIRCam Slitless Spectroscopy}}},
  author = {Sun, Fengwu and Egami, Eiichi and Pirzkal, Nor and Rieke, Marcia and Baum, Stefi and Boyer, Martha and Boyett, Kristan and Bunker, Andrew J. and Cameron, Alex J. and Curti, Mirko and Eisenstein, Daniel J. and Gennaro, Mario and Greene, Thomas P. and Jaffe, Daniel and Kelly, Doug and Koekemoer, Anton M. and Kumari, Nimisha and Maiolino, Roberto and Maseda, Michael and Perna, Michele and Rest, Armin and Robertson, Brant E. and Schlawin, Everett and Smit, Renske and Stansberry, John and Sunnquist, Ben and Tacchella, Sandro and Williams, Christina C. and Willmer, Christopher N. A.},
  year = 2023,
  month = aug,
  journal = {The Astrophysical Journal},
  volume = {953},
  pages = {53},
  publisher = {IOP},
  issn = {0004-637X},
  doi = {10.3847/1538-4357/acd53c},
  urldate = {2025-07-21}
}

@article{tepper-garcia2006,
  title = {Voigt Profile Fitting to Quasar Absorption Lines: An Analytic Approximation to the {{Voigt-Hjerting}} Function},
  shorttitle = {Voigt Profile Fitting to Quasar Absorption Lines},
  author = {{Tepper-Garc{\'i}a}, Thorsten},
  year = 2006,
  month = jul,
  journal = {Monthly Notices of the Royal Astronomical Society},
  volume = {369},
  pages = {2025--2035},
  publisher = {OUP},
  issn = {0035-8711},
  doi = {10.1111/j.1365-2966.2006.10450.x},
  urldate = {2024-11-05}
}

@article{trac2015,
  title = {{{SCORCH I}}: {{The Galaxy-Halo Connection}} in the {{First Billion Years}}},
  shorttitle = {{{SCORCH I}}},
  author = {Trac, Hy and Cen, Renyue and Mansfield, Philip},
  year = 2015,
  month = nov,
  journal = {The Astrophysical Journal},
  volume = {813},
  pages = {54},
  publisher = {IOP},
  issn = {0004-637X},
  doi = {10.1088/0004-637X/813/1/54},
  urldate = {2025-08-06}
}

@article{trenti2010,
  title = {The {{Galaxy Luminosity Function During}} the {{Reionization Epoch}}},
  author = {Trenti, M. and Stiavelli, M. and Bouwens, R. J. and Oesch, P. and Shull, J. M. and Illingworth, G. D. and Bradley, L. D. and Carollo, C. M.},
  year = 2010,
  month = may,
  journal = {The Astrophysical Journal},
  volume = {714},
  pages = {L202-L207},
  publisher = {IOP},
  issn = {0004-637X},
  doi = {10.1088/2041-8205/714/2/L202},
  urldate = {2025-07-31}
}

@article{vale2004,
  title = {Linking Halo Mass to Galaxy Luminosity},
  author = {Vale, A. and Ostriker, J. P.},
  year = 2004,
  month = sep,
  journal = {Monthly Notices of the Royal Astronomical Society},
  volume = {353},
  pages = {189--200},
  publisher = {OUP},
  issn = {0035-8711},
  doi = {10.1111/j.1365-2966.2004.08059.x},
  urldate = {2025-08-11}
}

@article{vandervelden2020,
  title = {{{CMasher}}: {{Scientific}} Colormaps for Making Accessible, Informative and 'cmashing' Plots},
  shorttitle = {{{CMasher}}},
  author = {{van der Velden}, Ellert},
  year = 2020,
  month = feb,
  journal = {The Journal of Open Source Software},
  volume = {5},
  pages = {2004},
  doi = {10.21105/joss.02004},
  urldate = {2022-06-07}
}

@article{viel2004,
  title = {Inferring the Dark Matter Power Spectrum from the {{Lyman}} {$\alpha$} Forest in High-Resolution {{QSO}} Absorption Spectra},
  author = {Viel, Matteo and Haehnelt, Martin G. and Springel, Volker},
  year = 2004,
  month = nov,
  journal = {Monthly Notices of the Royal Astronomical Society},
  volume = {354},
  pages = {684--694},
  publisher = {OUP},
  issn = {0035-8711},
  doi = {10.1111/j.1365-2966.2004.08224.x},
  urldate = {2024-11-05}
}

@article{virtanen2020,
  title = {{{SciPy}} 1.0: Fundamental Algorithms for Scientific Computing in {{Python}}},
  shorttitle = {{{SciPy}} 1.0},
  author = {Virtanen, Pauli and Gommers, Ralf and Oliphant, Travis E. and Haberland, Matt and Reddy, Tyler and Cournapeau, David and Burovski, Evgeni and Peterson, Pearu and Weckesser, Warren and Bright, Jonathan and {van der Walt}, St{\'e}fan J. and Brett, Matthew and Wilson, Joshua and Millman, K. Jarrod and Mayorov, Nikolay and Nelson, Andrew R. J. and Jones, Eric and Kern, Robert and Larson, Eric and Carey, C. J. and Polat, {\.I}lhan and Feng, Yu and Moore, Eric W. and VanderPlas, Jake and Laxalde, Denis and Perktold, Josef and Cimrman, Robert and Henriksen, Ian and Quintero, E. A. and Harris, Charles R. and Archibald, Anne M. and Ribeiro, Ant{\^o}nio H. and Pedregosa, Fabian and {van Mulbregt}, Paul and {SciPy 1. 0 Contributors}},
  year = 2020,
  month = feb,
  journal = {Nature Methods},
  volume = {17},
  pages = {261--272},
  doi = {10.1038/s41592-019-0686-2},
  urldate = {2022-06-07}
}

@article{wang2023,
  title = {A {{SPectroscopic Survey}} of {{Biased Halos}} in the {{Reionization Era}} ({{ASPIRE}}): {{JWST Reveals}} a {{Filamentary Structure}} around a z = 6.61 {{Quasar}}},
  shorttitle = {A {{SPectroscopic Survey}} of {{Biased Halos}} in the {{Reionization Era}} ({{ASPIRE}})},
  author = {Wang, Feige and Yang, Jinyi and Hennawi, Joseph F. and Fan, Xiaohui and Sun, Fengwu and Champagne, Jaclyn B. and Costa, Tiago and Habouzit, Melanie and Endsley, Ryan and Li, Zihao and Lin, Xiaojing and Meyer, Romain A. and Schindler, Jan--Torge and Wu, Yunjing and Ba{\~n}ados, Eduardo and Barth, Aaron J. and Bhowmick, Aklant K. and Bieri, Rebekka and Blecha, Laura and Bosman, Sarah and Cai, Zheng and Colina, Luis and Connor, Thomas and Davies, Frederick B. and Decarli, Roberto and De Rosa, Gisella and Drake, Alyssa B. and Egami, Eiichi and Eilers, Anna-Christina and Evans, Analis E. and Farina, Emanuele Paolo and Haiman, Zoltan and Jiang, Linhua and Jin, Xiangyu and Jun, Hyunsung D. and Kakiichi, Koki and Khusanova, Yana and Kulkarni, Girish and Li, Mingyu and Liu, Weizhe and Loiacono, Federica and Lupi, Alessandro and Mazzucchelli, Chiara and Onoue, Masafusa and Pudoka, Maria A. and {Rojas-Ruiz}, Sof{\'i}a and Shen, Yue and Strauss, Michael A. and Tee, Wei Leong and Trakhtenbrot, Benny and Trebitsch, Maxime and Venemans, Bram and Volonteri, Marta and Walter, Fabian and Xie, Zhang-Liang and Yue, Minghao and Zhang, Haowen and Zhang, Huanian and Zou, Siwei},
  year = 2023,
  month = jul,
  journal = {The Astrophysical Journal Letters},
  volume = {951},
  number = {1},
  pages = {L4},
  issn = {2041-8205, 2041-8213},
  doi = {10.3847/2041-8213/accd6f},
  urldate = {2025-11-28},
  langid = {english}
}

@article{weinberger2019,
  title = {Modelling the Observed Luminosity Function and Clustering Evolution of {{Ly}} {$\alpha$} Emitters: Growing Evidence for Late Reionization},
  shorttitle = {Modelling the Observed Luminosity Function and Clustering Evolution of {{Ly}} {$\alpha$} Emitters},
  author = {Weinberger, Lewis H. and Haehnelt, Martin G. and Kulkarni, Girish},
  year = 2019,
  month = may,
  journal = {Monthly Notices of the Royal Astronomical Society},
  volume = {485},
  pages = {1350--1366},
  publisher = {OUP},
  issn = {0035-8711},
  doi = {10.1093/mnras/stz481},
  urldate = {2025-08-11}
}

@article{wilkins2023,
  title = {First Light and Reionization Epoch Simulations ({{FLARES}}) {{XI}}: [{{O III}}] Emitting Galaxies at 5 {$<$} z {$<$} 10},
  shorttitle = {First Light and Reionization Epoch Simulations ({{FLARES}}) {{XI}}},
  author = {Wilkins, Stephen M. and Lovell, Christopher C. and Vijayan, Aswin P. and Irodotou, Dimitrios and Adams, Nathan J. and Roper, William J. and Caruana, Joseph and Matthee, Jorryt and Seeyave, Louise T. C. and Conselice, Christopher J. and {P{\'e}rez-Gonz{\'a}lez}, Pablo G. and Turner, Jack C. and Donnellan, James M. S. and Verma, Aprajita and Trussler, J. A. A.},
  year = 2023,
  month = jul,
  journal = {Monthly Notices of the Royal Astronomical Society},
  volume = {522},
  pages = {4014--4027},
  publisher = {OUP},
  issn = {0035-8711},
  doi = {10.1093/mnras/stad1126},
  urldate = {2025-11-13}
}

@article{zhu2023,
  title = {Probing {{Ultralate Reionization}}: {{Direct Measurements}} of the {{Mean Free Path}} over 5 {$<$} z {$<$} 6},
  shorttitle = {Probing {{Ultralate Reionization}}},
  author = {Zhu, Yongda and Becker, George D. and Christenson, Holly M. and D'Aloisio, Anson and Bosman, Sarah E. I. and Bakx, Tom and D'Odorico, Valentina and Bischetti, Manuela and Cain, Christopher and Davies, Frederick B. and Davies, Rebecca L. and Eilers, Anna-Christina and Fan, Xiaohui and Gaikwad, Prakash and Haehnelt, Martin G. and Keating, Laura C. and Kulkarni, Girish and Lai, Samuel and Ma, Hai-Xia and Mesinger, Andrei and Qin, Yuxiang and Satyavolu, Sindhu and Takeuchi, Tsutomu T. and Umehata, Hideki and Yang, Jinyi},
  year = 2023,
  month = oct,
  journal = {The Astrophysical Journal},
  volume = {955},
  pages = {115},
  publisher = {IOP},
  issn = {0004-637X},
  doi = {10.3847/1538-4357/aceef4},
  urldate = {2024-11-08}
}

@article{zhu2024,
  title = {On the {{Physical Nature}} of {{Ly$\alpha$ Transmission Spikes}} in {{High-redshift Quasar Spectra}}},
  author = {Zhu, Hanjue and Gnedin, Nickolay Y. and Avestruz, Camille},
  year = 2024,
  month = nov,
  journal = {The Astrophysical Journal},
  volume = {975},
  pages = {115},
  publisher = {IOP},
  issn = {0004-637X},
  doi = {10.3847/1538-4357/ad793c},
  urldate = {2024-11-11}
}

@article{zhu2026,
  title = {Galaxy {{Underdensities Host}} the {{Clearest Intergalactic Medium Ly$\alpha$ Transmission}} and {{Indicate Anisotropic Reionization}}},
  author = {Zhu, Yongda and Becker, George D. and D'Aloisio, Anson and Endsley, Ryan and Gangolli, Nakul and Cain, Christopher and Mason, Charlotte A. and Hashemi, Seyedazim and Hong, Hui},
  year = 2026,
  month = may,
  journal = {The Astrophysical Journal},
  volume = {1002},
  pages = {93},
  publisher = {IOP},
  issn = {0004-637X},
  doi = {10.3847/1538-4357/ae5bbb},
  urldate = {2026-05-28}
}
\bibliographystyle{aasjournal}

\appendix

\section{Impact of duty cycle timescale}
\label{sec:impact-duty-cycle}
In this appendix we detail the impact that changing the duty cycle,
defined in \eref{eq:duty-cycle}, has on the projected autocorrelation
function $\chi_V$ and the galaxy--\lya{} transmission cross-correlation
$\delta_F$. We explore three additional duty cycles, $\Delta t=25$, 100 and
200~Myr.

\begin{figure}
  \centering
  \includegraphics[width=\columnwidth]{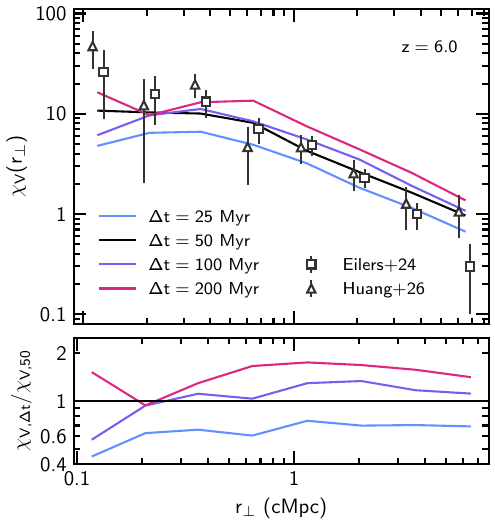}
  \caption{Projected autocorrelation function $\chi_V$ of \oiii{}
    emitters at $z = 6$ (top panel) for our fiducial duty cycle
    $\Delta t=50~{\rm Myr}$ (black), as well as for duty cycles of
    $\Delta t = 25$, 100 and 200 Myr (blue, purple and pink, respectively),
    along with observational constraints due to \citet{eilers2024}
    (grey squares) and \citet{huang2026} (grey triangles). We also
    show the ratio of $\chi_V$ for each of these duty cycles to
    $\chi_V$ for our fiducial choice of $\Delta t = 50~{\rm Myr}$ (bottom
    panel).}
  \label{fig:duty-cycle-chi}
\end{figure}

In \fref{fig:duty-cycle-chi}, we show the projected autocorrelation
$\chi_V$ at $z=6$. Modulo scatter at small $r_\perp$, which is largely due to
small numbers of haloes at these separations, the effect of increasing
the duty cycle is to increase $\chi_V$ across all $r_\perp$. This is easily
understood because, at fixed galaxy luminosity, increasing
(decreasing) the duty cycle tends to increase (decrease) the mass of
halo hosting that galaxy. Larger mass haloes tend to cluster more
strongly, hence the behaviour of $\chi_V$ as a function of $\Delta t$. The
behaviour is similar for the entire range of redshifts that we
consider in this work.

\begin{figure}
  \centering
  \includegraphics[width=\columnwidth]{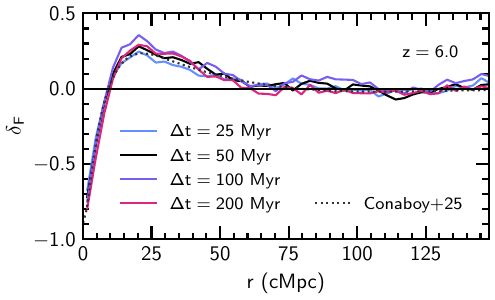}  
  \caption{Galaxy--\lya{} transmission correlation at $z=6$ for our
    fiducial duty cycle $\Delta t=50~{\rm Myr}$ (black), as well as for
    duty cycles of $\Delta t = 25$, 100 and 200 Myr (blue, purple and pink,
    respectively), along with the distribution from
    \citet{conaboy2025} (grey dotted).}
  \label{fig:duty-cycle-delta-F}
\end{figure}

\begin{figure}
  \centering
  \includegraphics[width=\columnwidth]{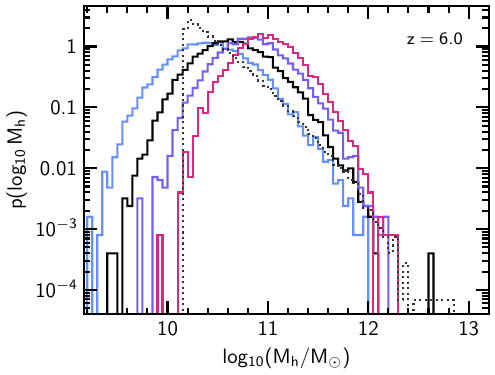}
  \caption{Distribution of halo masses used to compute $\delta_F$ at
    $z=6$ for our fiducial duty cycle $\Delta t=50~{\rm Myr}$ (black), as
    well as for duty cycles of $\Delta t = 25$, 100 and 200 Myr (blue,
    purple and pink, respectively), along with the distribution from
    \citet{conaboy2025} (grey dotted). The line styles are as in
    \fref{fig:duty-cycle-delta-F}.}
  \label{fig:lMh-delta-F}
\end{figure}

In \fref{fig:duty-cycle-delta-F} we show the galaxy--\lya{}
transmission correlation $\delta_F$ for these additional duty cycle
timescales. We have used a bin size four times smaller than that used
in \fref{fig:delta_F} to highlight the differences between the
different duty cycle timescales at small $r$. The differences between
the different duty cycle timescales are very small, and manifest
mostly at $r \lesssim 10~{\rm cMpc}$, where a longer duty cycle timescale
tends to decrease $\delta_F$. As described earlier in this appendix, and
demonstrated in \fref{fig:lMh-delta-F}, where we show the distribution
of halo masses used in the calculation of $\delta_F$ for each duty cycle
timescale $\Delta t$, increasing the duty cycle tends to increase the
typical halo mass for a given galaxy luminosity. As shown in
\citet{conaboy2025}, this leads to a decrease in $\delta_F$ at
$r \lesssim 10~{\rm cMpc}$. \citet{conaboy2025} also showed that the
behaviour at larger $r$ is complicated by the competing effects of
density and photoionisation rate enhancement around more massive
galaxies, and in this work we find that the scatter between different
timescales means that we observe no clear dependence on $\Delta t$ for
$r \gtrsim 10~{\rm cMpc}$. \fref{fig:lMh-dist} shows that the typical halo
mass used to compute $\delta_F$ differs by about $\sim 0.8~{\rm dex}$ between
the shortest and longest timescales, and that the distribution of
masses covers similar ranges. This explains why the difference in
$\delta_F$ for the different timescales is small.

\section{Impact of halo finder on small-scale clustering}
\label{sec:one-halo-term}
\begin{figure}
  \centering
  \includegraphics[width=\linewidth]{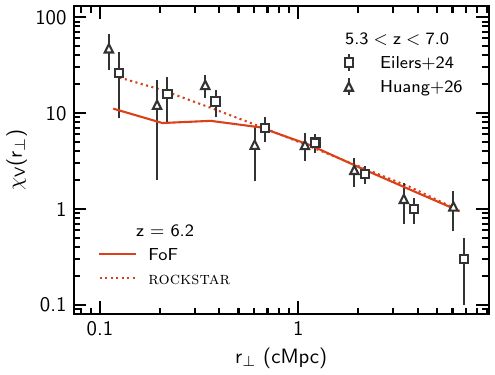}
  \caption{Comparison of $\chi_V$ at $z=6.2$ estimated using our fiducial
    FoF halo catalogue (solid line) and using a {\sc rockstar}
    catalogue (dotted line). Also shown are observational constraints
    due to \citet{eilers2024} (grey squares) and \citet{huang2026}
    (grey triangles).}
  \label{fig:one-halo-term}
\end{figure}

In this appendix we discuss the impact of our choice of halo finder on
the small-scale clustering of galaxies. In \fref{fig:one-halo-term} we
show the volume-averaged projected autocorrelation function $\chi_V$ (see
\sref{sec:galaxy-clustering}) at $z=6.2$, using our fiducial
abundance-matched FoF halo catalogue (taking the fiducial luminosity
cut of $\log_{10}(L_{\rm \oiii}/{\rm erg\,s^{-1}})>42.4$) and also for
a reanalysis of the snapshot using {\sc rockstar}
\citep{behroozi2013}. For the {\sc rockstar} case, we include all
(sub)haloes more massive than $\log_{10}(M_h/\msolh) > 10.5$, where
this lower mass limit is chosen to reproduce our abundance-matched
$\chi_V$ for $r_\perp>1~{\rm cMpc}$. We find that, for
$r_\perp \lesssim 0.4~{\rm cMpc}$, where the slope of $\chi_V$ for our fiducial
catalogue flattens, the {\sc rockstar} catalogue is in good agreement
with observational constraints. As described in
\sref{sec:galaxy-clustering}, {\sc rockstar} -- unlike FoF used with a
single linking length -- provides information about substructure and
hence on galaxies that can occupy the same halo. We can therefore
attribute the poorer agreement between our fiducial catalogue and the
observational data at small projected distances to this missing
contribution from galaxies inside the same halo. As discussed in
\sref{sec:galaxy-clustering}, we do not use the {\sc rockstar}
catalogue for our main analysis, since the underlying radiative
transfer simulation used the FoF catalogue to determine the location
of ionising sources, and so changing catalogues would break the
spatial correspondence between halo and source location.

\section{\oiii{}--UV ratio fitting}
\label{sec:oiii-uv-fit}
\begin{figure}
  \centering
\includegraphics[width=\columnwidth]{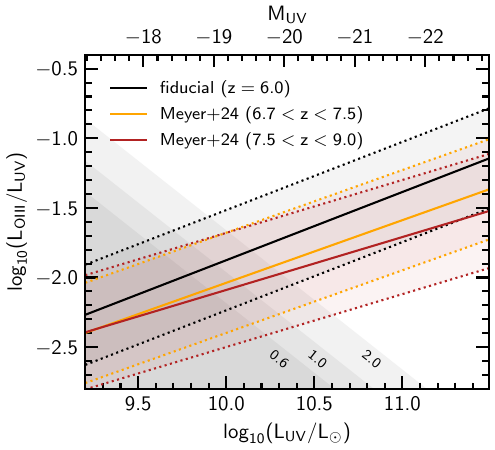}  
\caption{Mean \oiiia{}--UV relation (solid lines) and scatter (shaded
  band, dotted lines) as a function of $L_{\rm UV}$ for our fiducial
  model with
  $f_{\rm lim}=1\times 10^{-18}~{\rm erg}\,{\rm s^{-1}}\,{\rm cm^{-2}}$
  (black). Also shown is the relation inferred by \citet{meyer2024}
  (red and orange), whose methodology we follow. Also shown as grey
  shaded regions are ratioes excluded for
  $f_{\rm lim}/10^{-18}~{\rm erg}\,{\rm s^{-1}}\,{\rm cm^{-2}}=0.6,$
  1.0 and 2.0 corresponding to the $3\sigma$ limit for the EIGER programme
  \citepalias{matthee2023}, our fiducial limit, and the $5\sigma$ limit for
  the ASPIRE programme \citep{wang2023}, respectively.
  \label{fig:oiii-uv-ratio}}
\end{figure}

In \fref{fig:oiii-uv-ratio} we show the \oiiia--UV ratio for our
fiducial model (carrying out the fitting at $z=6$ assuming
$f_{\rm lim}=1\times 10^{-18}~{\rm erg\, s^{-1}\, cm^{-2}}$) compared to
the relation inferred by \citet{meyer2024} (in order to fit the
observed decline of \oiii-selected UV-bright galaxies), whose
methodology we follow in \sref{sec:uv-mag}. We find that our relation
is consistent with those found by \citet{meyer2024}, which are fit at
higher redshifts ($z\approx7.1$ and 7.9).

\vspace{1em}

\end{document}